%% file: waterWavePatches.tex
\documentclass[a4paper,12pt]{article}

\title{Large-scale simulation of shallow water waves with computation only on small staggered patches}

\author{
J.E. Bunder
\thanks{School of Mathematical Sciences,
University of Adelaide, South Australia.}
\thanks{\protect\url{http://orcid.org/0000-0001-5355-2288}}
\and
J. Divahar
\footnotemark[1]
\thanks{\protect\url{https://orcid.org/0000-0002-9506-8846}}
\and
Ioannis G. Kevrekidis
\thanks{Departments of Chemical and Biomolecular Engineering \& Applied Mathematics and Statistics, Johns Hopkins University, Baltimore, Maryland, USA.
\protect\url{https://orcid.org/0000-0003-2220-3522}}
\and
Trent W. Mattner
\footnotemark[1]
\thanks{\protect\url{https://orcid.org/0000-0002-5313-5887}}
\and
A.J. Roberts
\footnotemark[1]
\thanks{\protect\url{http://orcid.org/0000-0001-8930-1552},
\protect\url{mailto:anthony.roberts@adelaide.edu.au}}
}

\usepackage{natbib}
\AtEndDocument{\raggedright\bibliography{bibexport}}
\usepackage{amsmath,amssymb,microtype,defns,euler} 
\usepackage{graphicx} 
\def\raisedName#1{} 

\usepackage{listings}
\usepackage[capitalise,nameinlink,noabbrev]{cleveref}
\crefname{equation}{}{} 
\let\eqref\cref

\usepackage{pgfplots}
\pgfplotsset{compat=newest}
\usepgfplotslibrary{external}
\tikzsetexternalprefix{Figs/}

\usepackage{url,doi}

\def\i{\operatorname{i}}
\Bb R\Bb Z
\Vec k\Vec u
\Cal G\Cal J \Cal V\Cal L

\newcommand{\asinh}{\operatorname{arcsinh}}

\newcommand{\uu}{{\bar u}}
\newcommand{\vv}{{\bar v}}
\newcommand{\cc}{{\bar c}}
\newcommand{\bq}{{\bar q}}

\renewcommand{\vec}[1]{\text{\boldmath$#1$}}
\newcounter{i}
\newcounter{j}

\begin{document}

\maketitle

\begin{abstract}
The multiscale patch scheme is built from given small micro-scale simulations of complicated physical processes to empower large macro-scale simulations.  By coupling small patches of simulations over unsimulated spatial gaps, large savings in computational time are possible.  Here we discuss generalising the patch scheme to the case of wave systems on staggered grids in 2D space.  Classic macro-scale interpolation provides a generic coupling between patches that achieves arbitrarily high order consistency between the emergent macro-scale simulation and the underlying micro-scale dynamics.  Eigen-analysis indicates that the resultant scheme empowers feasible computation of large macro-scale simulations of wave systems even with complicated underlying physics.  As examples we use the scheme to simulate some wave scenarios via a turbulent shallow water model. 
\end{abstract}

\tableofcontents

\section{Introduction}

\begin{figure}
\centering\caption{\label{fig:bores} bores, such as occur in tsunamis and some floods, exhibit high levels of turbulence that should be accounted for by turbulent shallow water wave models (left, Qiantang River \protect\cite[Fig.~1(a)]{Reungoat2018}; right, by D.H. Peregrine \protect\cite[\#199]{VanDyke82}).}
\includegraphics[height=0.29\linewidth]{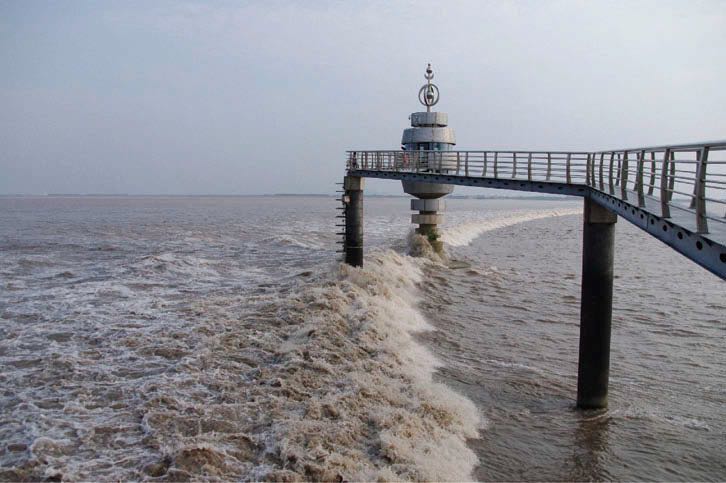}
\includegraphics[height=0.29\linewidth]{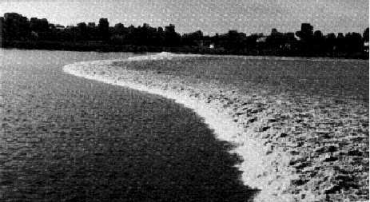}
\end{figure}

Partial differential equations describing rivers, bores, floods and tsunamis (\cref{fig:bores}) are typically written at the macro-scale of kilometres. 
But the level at which the underlying turbulent fluid physics are best understood is at the much finer sub-metre scale. 
Although extant macro-scale models for floods and tsunamis are well established, for many other multiscale and multiphysics wave-like problems good macro-scale descriptions (good closures) do not exist. 
We aim to empower scientists and engineers to use brief bursts of micro-scale wave-like simulation on small patches of the space-time domain in order to make efficient accurate macro-scale simulations without ever needing to know an explicit algebraic macro-scale closure.
Building upon the 1D~space techniques of \cite{Cao2014a}, here we develop, test and analyse techniques for waves in 2D~space.
Then future developments are planned to directly simulate metre-sized patches of turbulent flow, and couple these patches together over a hundred metres of unsimulated space in order to improve large scale flood and tsunami predictions.

Another area of application is in geophysical fluid dynamics where analogous issues arise in climate and weather models.
\cite{Grabowski2001} commented that ``cloud-related small-scale processes play important roles in large-scale atmospheric flows and are essential for both weather and climate. \ldots\ design an approach in which a 2D cloud-resolving model is applied in each column of the large-scale model to represent subgrid-scale cloud processes.''
\cite{Grooms2018} described the approach as ``the small-scale equations are solved on a set of many small-area domains \ldots, and a single high-resolution model of small-scale dynamics is embedded within each column of the large-scale model.''   
That is, relatively small patches of a high resolution cloud resolving model are coupled into a large scale computational model.  
The fundamental challenge we address is how to couple across 2D space such small scale patches of wave-like dynamics in order to accurately compute the macro-scale.

Many multiscale modelling techniques have been recently developed for dissipative systems~\cite[e.g.]{E:2003fk, Kevrekidis:2003fk, Roberts:2005uq, Hou:2008fk}.
Our macro-scopic modelling further develops the equation-free patch scheme \cite[e.g.]{Gear03, Samaey03b, Samaey:2005fk, Samaey2009}, also known as the gap-tooth scheme.
This paper specifically develops the scheme to simulate wave-like systems over large time and large multi-D space scales using some \emph{given} micro-scale simulator.
A major distinction with other work is that our scheme only simulates on small well-separated patches of space, whereas other multiscale approaches typically compute over all space: for example, see the numerical homogenization of \cite{Maier2019}, or the computational homogenization discussed by \cite{Geers2010}.

\begin{figure}
\begin{tabular}{@{}cc@{}}
\parbox[b]{0.47\linewidth}{(a) $t=0$\\
\includegraphics[width=\linewidth]{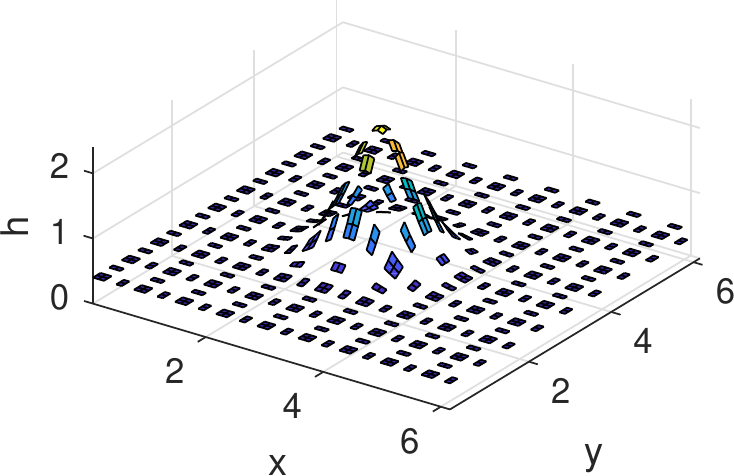}}
&
\parbox[b]{0.47\linewidth}{\caption{\label{fig:PatchHumpSim}indicative simulation of nonlinear shallow water shows the evolving water depth~\(h(x,y,t)\) at three times.  
Micro-scale computations are only done in `small' patches (for better visibility, here larger than necessary) of the spatial \(xy\)-domain; the hump at time zero slumps down and out over time.}}
\\
\parbox[b]{0.47\linewidth}{(b) $t=0.4$\\
\includegraphics[width=\linewidth]{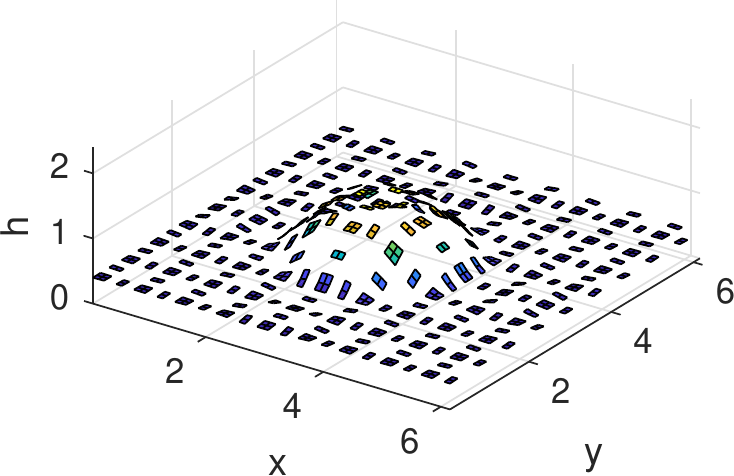}}
&
\parbox[b]{0.47\linewidth}{(c) $t=0.8$\\
\includegraphics[width=\linewidth]{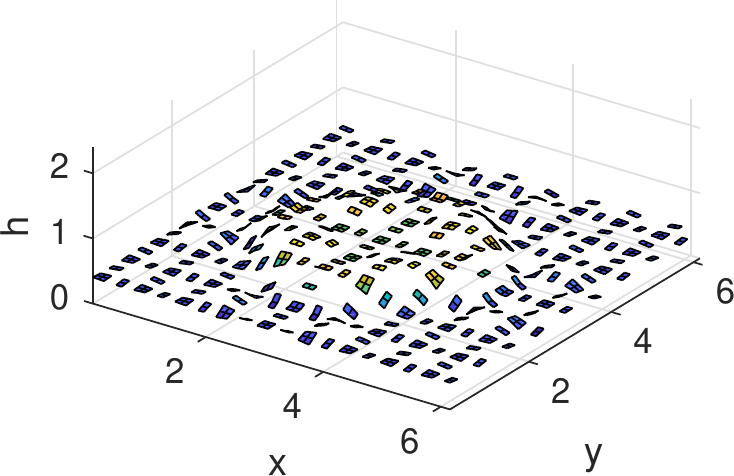}}
\end{tabular}
\end{figure}

As illustrated in the nonlinear shallow water wave simulation of \cref{fig:PatchHumpSim}, we divide space into distributed but relatively small patches within which we compute with the micro-scale simulator, and which are separated by unsimulated space (the patches of \cref{fig:PatchHumpSim} are larger than necessary for simulation solely in order to be visible).
The patch scheme models the macro-scale quantities over large space scales via coupling, over the unsimulated space, of the micro-scale simulations in each patch.
This scheme is designed for cases when the wave-like micro-scale simulator is computationally expensive so that only relatively small time and spatial domain simulations are feasible, such as turbulent floods and cloud physics.
In the scheme, the micro-scale simulator provides the necessary data for the macro-scopic model, so whenever the micro-scale simulator is refined by a modeller, then the overall macro-scale simulation correspondingly improves.

\cref{fig:PatchHumpSim} illustrates our patch scheme for 2D waves---a scheme that extends the equation-free methodology \cite[]{Kevrekidis09a, Roberts04d}.
By simulating the wave details only on small patches in a large spatial domain, we greatly reduce the expense of computation over the domain, and make feasible very large domain simulations of micro-scales.
The key challenge is the following: how do we couple patches of wave computation?
This article answers by showing that straightforward low-order polynomial interpolation works wonderfully, and a spectral coupling is more accurate (\cref{sec:spsm2d}).

The patch method and our theoretical support (\cref{sec:spsm2d,sec:ildamv}) adapts to whatever micro-scale simulator is provided. 
\cref{sec:ntfs} describes how to implement the patch method by coupling small patches of the Smagorinski micro-scale dynamics to simulate floods over a macro-scale space.
The analysis of \cref{sec:ecmclc} indicates that the patches can occupy as small a fraction of space as is necessary for a good micro-scale simulation without affecting macro-scale accuracy, thus indicating large computational gains are feasible with the methodology.

\begin{figure}
\tikzsetfigurename{scalesep}
\begin{tabular}{@{}cc@{}}
\begin{tikzpicture}
  \begin{axis}[xlabel={$k_x$},ylabel={$k_y$},zlabel={$\omega$}
  ,domain=-1.99:1.99
  ,xmin=-2,xmax=2,ymin=-2,ymax=2,zmin=0,zmax=2
  ,view={30}{60},grid,width=0.5\linewidth
  ,ztick={0}
  ,xtick={-1,-0.3,0.3,1},
  ,xticklabels={$-\frac\pi d$,$-\frac\pi D$,$\frac\pi D$,$\frac\pi d$}
  ,ytick={-1,-0.3,0.3,1},
  ,yticklabels={$-\frac\pi d$,$-\frac\pi D$,$\phantom{+}\frac\pi D$,$\phantom{+}\frac\pi d$}
  ,title={(a) water waves: ideal}
  ]
  \foreach \s in {1} {
  \addplot3[surf,opacity=0.5]{\s*sqrt(sqrt(x^2+y^2)*tanh(sqrt(x^2+y^2))) };
  };
  \end{axis}
\end{tikzpicture} 
&
\parbox[b]{0.5\linewidth}{%
\caption{\label{fig:scalesep}The dispersion relation, frequency~$\omega$ versus  wavenumber~\kv\ (colour\({}\propto\omega\)), illustrates our macro-scale and micro-scale separation in wave systems: (a)~\(\omega=\sqrt{g|\kv|\tanh(|\kv|h)}\) for ideal water waves in depth~\(h\); (b)~represents the slow macro-scale waves in the common shallow water approximation, \(\omega=\sqrt{gh}|\kv|\); and (c)~schematically illustrates the gap between the slow macro-scale waves (centre) and fast, micro-scale, sub-patch waves (crinkly) that are also resolved by the patch scheme.}}
\\
\begin{tikzpicture}
  \begin{axis}[xlabel={$k_x$},ylabel={$k_y$},zlabel={$\omega$}
  ,domain=-0.33:0.33
  ,xmin=-2,xmax=2,ymin=-2,ymax=2,zmin=0,zmax=2
  ,view={30}{60},grid,width=0.5\linewidth
  ,ztick={0}
  ,xtick={-1,-0.3,0.3,1},
  ,xticklabels={$-\frac\pi d$,$-\frac\pi D$,$\frac\pi D$,$\frac\pi d$}
  ,ytick={-1,-0.3,0.3,1},
  ,yticklabels={$-\frac\pi d$,$-\frac\pi D$,$\phantom{+}\frac\pi D$,$\phantom{+}\frac\pi d$}
  ,title={(b) shallow: macro-scale \textsc{pde}}
  ]
  \foreach \s in {1} {
  \addplot3[surf,opacity=0.5,samples=7,point meta max=2]
    {\s*sqrt(sqrt(x^2+y^2)*tanh(sqrt(x^2+y^2))) };
  };
  \end{axis}
\end{tikzpicture} 
&
\begin{tikzpicture}
  \begin{axis}[xlabel={$k_x$},ylabel={$k_y$},zlabel={$\omega$}
  ,domain=-1.99:1.99
  ,xmin=-2,xmax=2,ymin=-2,ymax=2,zmin=0,zmax=2
  ,view={30}{60},grid,width=0.5\linewidth
  ,ztick={0}
  ,xtick={-1,-0.3,0.3,1},
  ,xticklabels={$-\frac\pi d$,$-\frac\pi D$,$\frac\pi D$,$\frac\pi d$}
  ,ytick={-1,-0.3,0.3,1},
  ,yticklabels={$-\frac\pi d$,$-\frac\pi D$,$\phantom{+}\frac\pi D$,$\phantom{+}\frac\pi d$}
  ,title={(c) macro/micro-scale patches}
  ]
  \foreach \s in {1} {
  \addplot3[surf,opacity=0.5,unbounded coords=jump,samples=35]
  {\s*sqrt(sqrt(x^2+y^2+0.0001)*tanh(sqrt(x^2+y^2+0.0001))) 
    +0.1*rand*(max(abs(x),abs(y))>1) 
    +inf*(abs(max(abs(x),abs(y))-0.7)<0.33)};
  };
  \end{axis}
\end{tikzpicture} 
\end{tabular}
\end{figure}
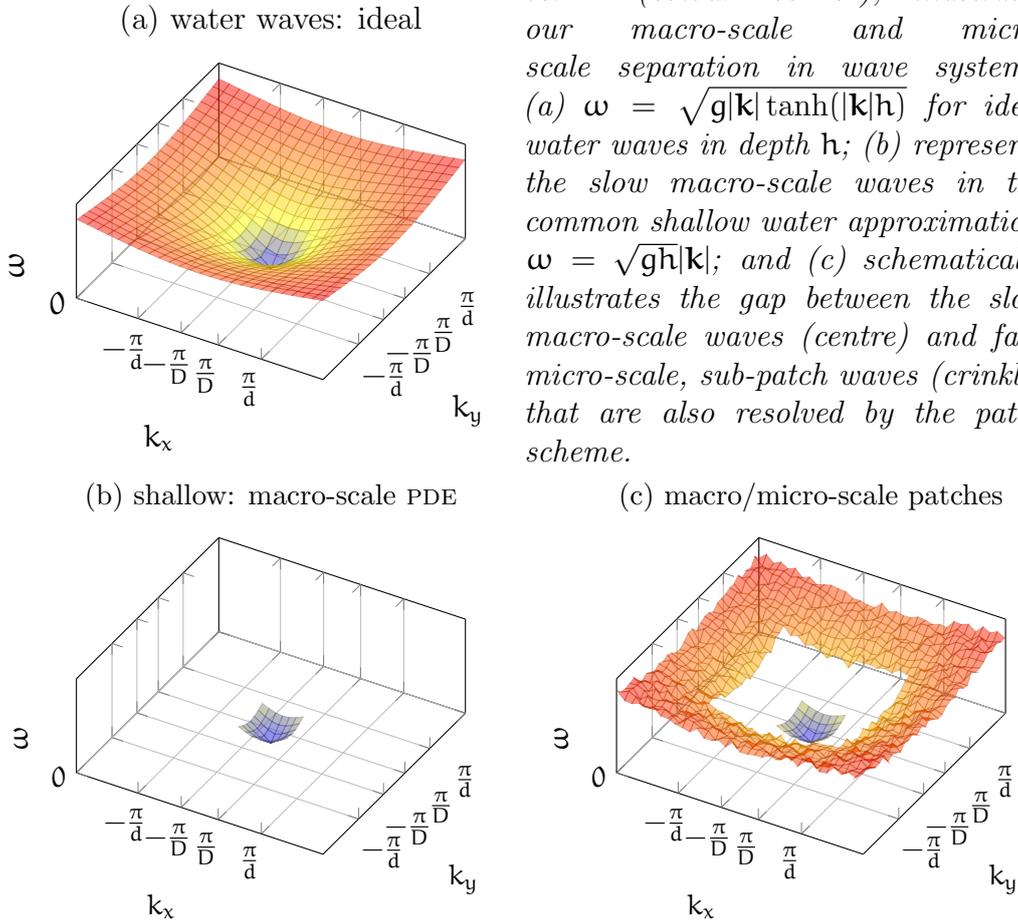

Multiscale methodologies and their analysis typically rely on a spectral gap in the eigenvalues of the system's dynamics, and the same holds here for our analysis of wave-like systems.
But importantly, the gap exists only due to our interest in long waves in some multi-physics system, in the scenario where the system is known only via micro-scale simulation.
Ideal wave-like systems typically do not have a spectral gap, as illustrated by \cref{fig:scalesep}(a) for ideal finite-depth 2D water waves. 
But to simulate tides, floods and tsunamis, researchers, physicists and engineers typically focus on the macro-scale dynamics of long waves, of wavelength significantly larger than the depth~\(h\), via a shallow water approximation, as illustrated in \cref{fig:scalesep}(b). 
The patch scheme for wave systems is designed to reproduce accurately such macro-scale long wave dynamics via simulation on patches, of small size\({}\propto d\).  
So there exists a spectral gap between the long waves and the micro-scale sub-patch waves as shown schematically in \cref{fig:scalesep}(c).
The spectral gap that we invoke arises only in our multiscale patch computational model of the system (\cref{fig:scalesep}(c)), and not (necessarily) in the physical system itself (\cref{fig:scalesep}(a)).

\section{Stagger patches of staggered microcode in 2D space}
\label{sec:spsm2d}

Often, wave-like systems in multi-D are written in terms of two conjugate variables. 
For example, conjugate variables may be position and momentum density, electric and magnetic fields, or water depth~\(h\) and mean lateral velocity~\(\uv=(u,v)\) as herein.
Such variables depend upon time and 2D~space \((x,y,t)\) (\cref{fig:PatchHumpSim}).
This article is phrased in the context of shallow water waves, but it applies to any wave-like system in the generic canonical 2D wave \pde{}s for `depth'~\(h(x,y,t)\) and `velocities' \(u(x,y,t)\) and~\(v(x,y,t)\): 
\begin{align}
&\D th=-\D xu-\D yv+\cdots;
&&\D tu=-\D xh +\cdots;
&&\D tv=-\D yh +\cdots.
\label{eqs:wavepdes}
\end{align}
In application the \pde{}s~\cref{eqs:wavepdes} would have various multiplicative constants in the right-hand sides: such constants do not materially change our analysis.
However, the ellipses,~\(\cdots\), in the right-hand sides~\cref{eqs:wavepdes} represent application specific nonlinear and/or higher derivative terms, neglected in our initial linear exploration,  but which form a significant complication to be addressed by \cref{sec:ildamv,sec:ntfs} in the context of nonlinear water waves.

This section generalises to 2D the staggered patches in 1D space that \cite{Cao2014a} developed to accurately simulate complex 1D wave propagation.
Researchers may easily implement for themselves 1D staggered patches for their own 1D wave-like problems using a \textsc{Matlab}\slash Octave Toolbox currently in development \cite[]{Roberts2019b}.
We plan that the Toolbox should soon support the 2D staggered patches developed herein.

\subsection{The staggered microgrid computational scheme}
\label{sec:smcs}

In principle the patch scheme uses a pre-existing computational model as the micro-scale simulator.
Here we suppose the computational model is a micro-scale discretisation of the canonical wave \pde{}s~\cref{eqs:wavepdes}, linearised.
It is well established that a staggered grid is a good way to discretise such linear wave \pde{}s.
Here let's create a 2D micro-scale staggered grid in the spatial \(xy\)-plane, of micro-scale spacing~\(d\), as shown below-left, where indices~\(i\) and~\(j\) step by one for each (green) microgrid line:
\footnote{Many people prefer that indices \(i,j\) increment by~\(1/2\) between each (green) microgrid line drawn, instead of the increment by one we choose. 
There are pros and cons with either choice.}
\begin{equation}
\input{microStaggerGrid}\quad
\begin{array}[b]{l@{\,}l}
\bullet&\displaystyle
\de t{h_{i,j}}=-\frac{u_{i+1,j}-u_{i-1,j}}{2d}-\frac{v_{i,j+1}-v_{i,j-1}}{2d}+\cdots;
\\[2ex]\color{cyan}\bullet&\displaystyle
\de t{u_{i,j}}=-\frac{h_{i+1,j}-h_{i-1,j}}{2d} +\cdots;
\\[2ex]\color{red}\bullet&\displaystyle
\de t{v_{i,j}}=-\frac{h_{i,j+1}-h_{i,j-1}}{2d} +\cdots.
\\\,
\end{array}
\label{eqs:waveds}
\end{equation}
At each of three-quarters of the 2D microgrid points, according to the coloured discs (i.e., one quarter for each of~$h$, $u$ and~$v$), we discretise the corresponding canonical terms of the wave \pde~\cref{eqs:wavepdes} with centred differences as indicated in the \ode{}s~\cref{eqs:waveds}.

The system of \ode{}s~\cref{eqs:waveds}, for the moment limited to the shown terms and completed with appropriate boundary and initial conditions, is our chosen canonical micro-scale simulator in 2D~space.
Alternative micro-scale systems could be developed using finite element, or finite volume methods \cite[e.g.]{LeVeque2011}, or a particle based method such as lattice Boltzmann \cite[e.g.]{Liu2009b}, molecular dynamics \cite[e.g.]{Southern2008}, or smoothed-particle hydrodynamics \cite[e.g.]{Monaghan92}.

A standard von Neumann stability analysis of~\cref{eqs:waveds} shows this micro-scale computational scheme is purely wave-like.
Ignoring boundaries, substitute \((h,u,v)\propto e^{\i(kx+\ell y)+\lambda t}\) into~\cref{eqs:waveds} on the microgrid (\(x_i=di\) and \(y_j=dj\), using the upright roman \(\i:=\sqrt{-1}\) as distinct from the microgrid index~\(i\) in math font), and then straightforward algebra leads to the eigen-problem
\begin{equation*}
\lambda d\begin{bmatrix} h\\u\\v \end{bmatrix}
=\begin{bmatrix} 0&-\i\sin kd&-\i\sin \ell d
\\-\i\sin kd&0&0
\\-\i\sin \ell d&0&0 \end{bmatrix}\begin{bmatrix} h\\u\\v \end{bmatrix}.
\end{equation*}
The corresponding characteristic equation, \(\lambda d\big[(\lambda d)^2+\sin^2kd+\sin^2\ell d\big]=0\), shows there are microgrid wave solutions with frequency \begin{equation*}
\omega=\pm\sqrt{\sin^2kd+\sin^2\ell d}\big/d\,.
\end{equation*}
Consequently, the microgrid~\cref{eqs:waveds} resolves waves with spatial wavenumbers \(|kd|,|\ell d|\leq\pi/2\), and does so with frequencies correct to errors~\Ord{k^4+\ell^4}.
The characteristic equation also shows that the staggered microgrid has neutral modes, \(\lambda d=0\), of vortical flow composed of modes where the ratio \(u:v=-\sin \ell d:\sin kd\) for every resolved wavenumber.

These waves and vortices have analogues in the next \cref{sec:smgp} on our 2D~staggered macrogrid of 2D~patches.

\subsection{Staggered macro-scale grid of patches}
\label{sec:smgp}

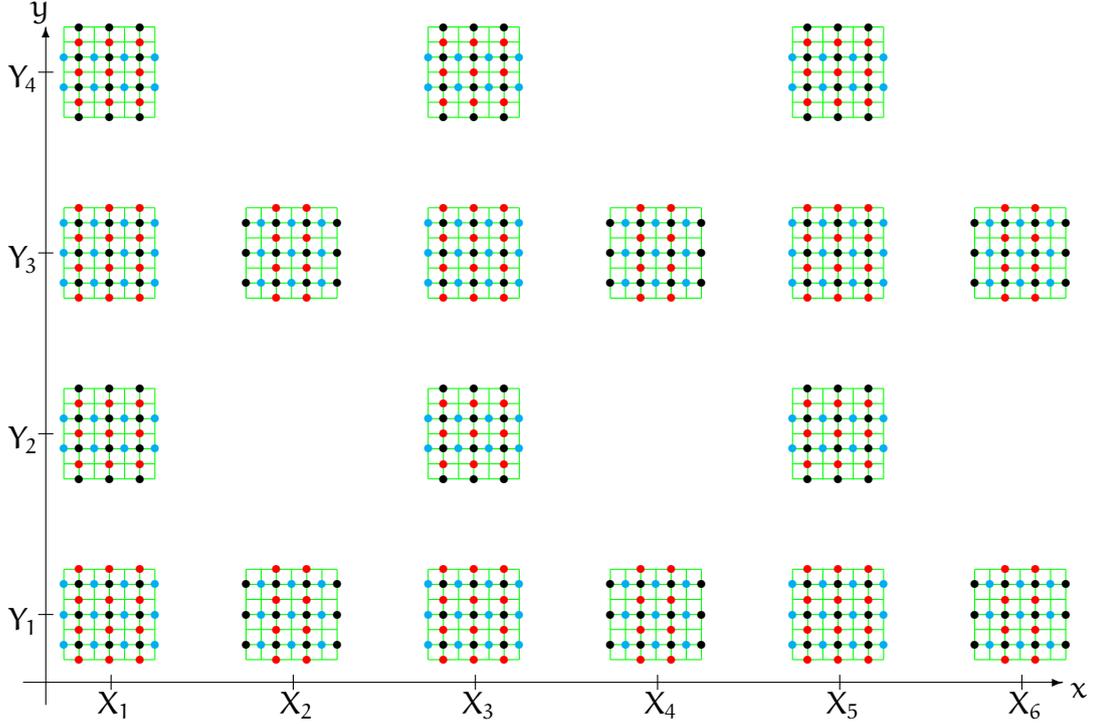
\begin{figure}
\centering\caption{\label{fig:patchMacro}2D staggered macro-scale grid of patches for the ideal wave system~\cref{eqs:waveds}, each patch a 2D staggered micro-scale grid (here \(7\times7\)).  
The centres,~\((X_I,Y_J)\), of each patch are spaced apart by the macro-scale length~\(D\).  
The three types of patches are characterised by the centre-value, as coloured: \({\color{black}\bullet}~H\)-patches;
\({\color{cyan}\bullet}~U\)-patches; and
\({\color{red}\bullet}~V\)-patches.}
\input{macroStaggerPatches}
\end{figure}

Because of the beautiful properties of a staggered grid for simulating waves, we choose the macro-scale grid of patches in space to be also staggered as illustrated in \cref{fig:patchMacro}.
One result of choosing this staggered macro-scale is that the resultant  macro-scale waves have an excellent frequency spectrum (\cref{fig:idealevals}).

In the analysis and simulations explored herein we seek to simulate waves in `large' rectangular domains~\(L_x\times L_y\), with periodic boundary conditions.
To form the multiscale grid of patches, we choose a macro-scale grid of points~\((X_I,Y_J)\) with constant macro-scale spacing~\(D\) in both directions (ongoing research is exploring different spacing), for integer indices~\(I=1,\ldots,N_x\) and~\(J=1,\ldots,N_y\).
Generally we use uppercase letters for macro-scale quantities, and lowercase for micro-scale, sub-patch, quantities (as in \cref{sec:smcs}).
Square patches are centred on this macro-scale grid (\cref{fig:patchMacro}) so that the micro-scale grid of the \((I,J)\)th~patch has
\begin{equation*}
x_i^{IJ}=X_I+id\,,\quad
y_j^{IJ}=Y_J+jd\,,\quad
i,j=-n,\ldots,n\,,
\end{equation*}
for some chosen odd~\(n\) (\(n=3\) in \cref{fig:patchMacro}).
The centre microgrid-point of each patch has micro-scale indices \(i=j=0\).

The macro-scale model and predictions are parametrised by the grid-value at the centre of each of the patches (as coloured in \cref{fig:patchMacro}):
\footnote{In problems with a heterogeneous micro-scale we would typically parametrise the macro-scale by so-called core-averages \cite[e.g.]{Bunder2013b}, but here there is no need.}
\begin{subequations}\label{eqs:macval}%
\begin{align}
\bullet\,&H_{IJ}(t):=h_{00}^{IJ}(t)\quad\text{for odd \(I\) and odd \(J\)};
\\\color{cyan}\bullet\,&U_{IJ}(t):=u_{00}^{IJ}(t)\quad\text{for even \(I\) and odd \(J\)};
\\\color{red}\bullet\,&V_{IJ}(t):=v_{00}^{IJ}(t)\quad\text{for odd \(I\) and even \(J\)};
\\&\text{and no patches for even \(I\) and even \(J\)}.
\end{align}
\end{subequations}
In these, and hereafter but only where needed, we denote micro-scale sub-patch quantities with superscripts~\(I,J\) to denote the particular patch.
Thus there are three types of patches making up the macro-scale grid of patches: they are called \(H\)-patch, \(U\)-patch, and \(V\)-patch corresponding to each patch's centre value.

\cref{fig:patchMacro} shows that these three different types of patches require a different combination of `boundary' values on the edges of each patch: we call these \emph{edge values} to distinguish them from macro-scale domain boundaries.
We couple the patches together by interpolating corresponding centre-patch values~\eqref{eqs:macval} to the edges of patches as needed.
The result is a multiscale computational model describing waves over the whole spatial domain (as in \cref{fig:PatchHumpSim}).

Every patch scheme has a key design parameter that measures the relative size of patches.
Here we use the non-dimensional ratio between the patch half-width~\(nd\) and the macro-scale patch spacing~\(D\):
\begin{equation}
r:=nd/D\,.
\label{eq:rat}
\end{equation}
\cref{fig:PatchHumpSim} shows a simulation with \(r=0.4\).
When this ratio \(r=0.5\) then the patches abut; when \(r=1\) the patches overlap as in `holistic discretisation' \cite[e.g.]{Roberts00a, Roberts2011a}.
In simulations we prefer small~\(r\) (small~\(n\)) in order to minimise computations: in 2D, for fixed micro-scale~\(d\), the amount of computation is~\(\propto r^2\).
For scenarios where the micro-scale is `smooth' \cref{sec:ecmclc} confirms that arbitrarily small~\(r\) may be used for wave systems, as previously established for dissipative systems \cite[e.g.]{Roberts00a, Roberts2011a}.
The three main constraints limiting the smallness of patches are:  the presence of any micro-scale heterogeneities \cite[e.g.]{Bunder2013b}; the stiffness of the multiscale patch scheme; and computational round-off error.

Here we discuss and analyse two specific interpolations that couple patches by giving patch-edge values: firstly, nearest-neighbour linear; and secondly, global spectral.
Ongoing research is exploring a variety of other 2D interpolation schemes to determine the patch-edge values.
In the scenario of very large scale computations of complex physics each patch may be allocated to each compute core (\cite{Lee2017} discussed fault tolerance with patches in exascale computation).
Then the nearest-neighbour linear interpolation is simple, gives acceptable basic accuracy (errors~\Ord{D^2}) and has the advantage of minimising inter-patch/core communication.
Spectral interpolation requires global synchronous communication, and a simple macro-scale domain, but is very accurate (errors exponentially small in~\(D\)).
The best approach will depend upon the complexity and size of the problem, and the nature of the utilised computer.

\subsubsection{Nearest-neighbour linear interpolation}
\label{ss:nnli}

There are three types of patches to consider, and two types of edge for each patch---see \cref{fig:patchMacro} throughout this discussion.
Each \({\color{black}\bullet}\,H\)-patch (odd~\(I\) and~\(J\)) requires \(u\)-values on the left and right edges, at \(x=X_I\pm nd=X_I\pm rD\). These are linearly interpolated as, constant in~\(j\) and recalling \(U_{I,J}=u_{0,0}^{I,J}\), 
\begin{subequations}\label{eqs:lint}%
\begin{equation}
{\color{black}\bullet}\ u^{I,J}_{\pm n,j}:=\tfrac12(1\pm r)U_{I+1,J}+\tfrac12(1\mp r)U_{I-1,J}\,.
\end{equation}
Correspondingly, each \({\color{black}\bullet}\,H\)-patch requires \(v\)-values on the top and bottom edges, at \(y=Y_J\pm nd=Y_J\pm rD\), computed as, constant in~\(i\), 
\begin{equation}
{\color{black}\bullet}\ v^{I,J}_{i,\pm n}:=\tfrac12(1\pm r)V_{I,J+1}+\tfrac12(1\mp r)V_{I,J-1}\,.
\end{equation}
Likewise, each \({\color{cyan}\bullet}\,U\)-patch (even~\(I\), odd~\(J\)) requires \(h\)-values on the left and right, whereas each \({\color{red}\bullet}\,V\)-patch (odd~\(I\), even~\(J\)) requires \(h\)-values on the top and bottom: respectively,
\begin{align}
{\color{cyan}\bullet}\ h^{I,J}_{\pm n,j}&:=\tfrac12(1\pm r)H_{I+1,J}+\tfrac12(1\mp r)H_{I-1,J}
\,,\\
{\color{red}\bullet}\ h^{I,J}_{i,\pm n}&:=\tfrac12(1\pm r)H_{I,J+1}+\tfrac12(1\mp r)H_{I,J-1}
\,.
\end{align}
Further (\cref{fig:patchMacro}), each \({\color{cyan}\bullet}\,U\)-patch requires \(v\)-values top and bottom, whereas each \({\color{red}\bullet}\,V\)-patch requires \(u\)-values left and right.
For the \((I,J)\)th~patch, obtain these via interpolation from the four diagonal nearest-neighbours, patches~\((I\pm1,J\pm1)\):
\begin{align}
{\color{cyan}\bullet}\ v^{I,J}_{i,\pm n}&:=\frac{1\pm r}4(V_{I+1,J+1}+V_{I-1,J+1})
+\frac{1\mp r}4(V_{I+1,J-1}+V_{I-1,J-1});
\\
{\color{red}\bullet}\ u^{I,J}_{\pm n,j}&:=\frac{1\pm r}4(U_{I+1,J+1}+U_{I+1,J-1})
+\frac{1\mp r}4(U_{I-1,J+1}+U_{I-1,J-1}).
\end{align}
\end{subequations}
More interpolated patch-edge values are required when implementing `viscous' dissipation (\cref{sec:ildamv}) or nonlinear gradients (\cref{sec:ntfs}).

\begin{figure}
\caption{\label{fig:lint}an example ideal wave propagation using patches (size ratio \(r=0.3\)) coupled by nearest-neighbour linear interpolation.  Plot the micro-scale \(h\)-field within each patch at the initial and three subsequent times covering nearly one wave period.}
\centering
\begin{tabular}{@{}cc@{}}
\parbox[b]{0.47\linewidth}{(a) $t=0$\\
\includegraphics[width=\linewidth]{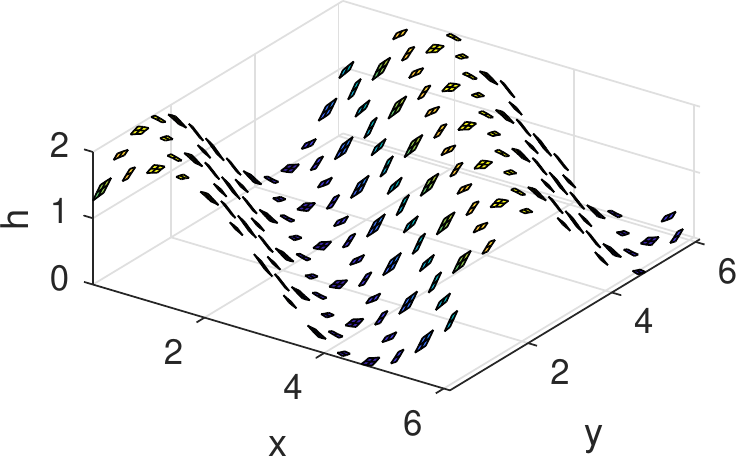}}
&
\parbox[b]{0.47\linewidth}{(b) $t=1.21$\\[1ex]
\includegraphics[width=\linewidth]{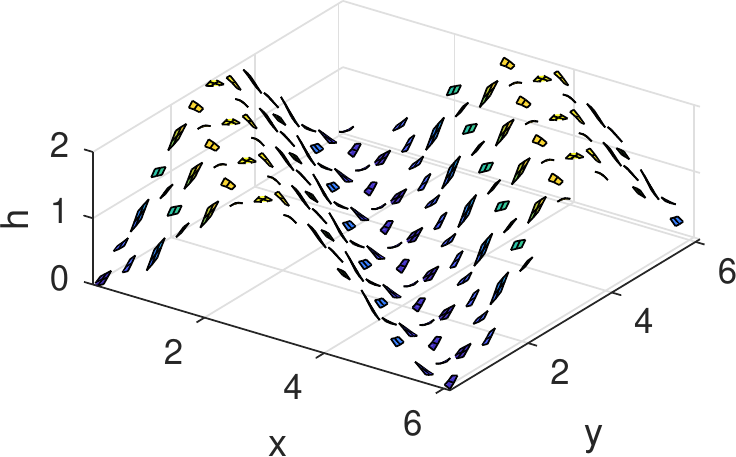}}
\\
\parbox[b]{0.47\linewidth}{(c) $t=2.42$\\
\includegraphics[width=\linewidth]{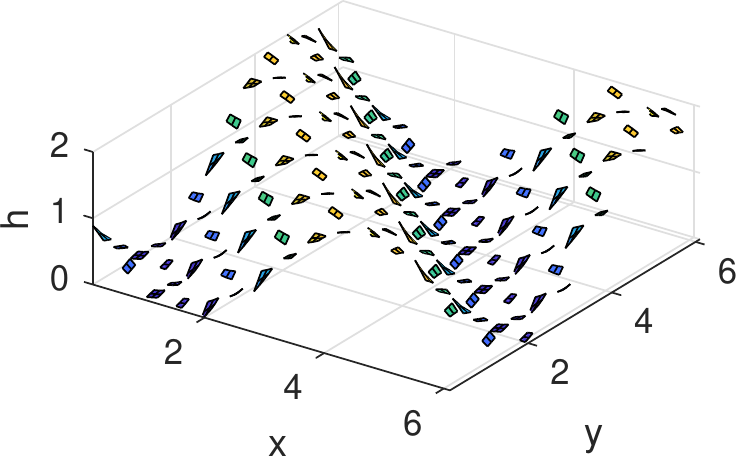}}
&
\parbox[b]{0.47\linewidth}{(d) $t=3.62$\\
\includegraphics[width=\linewidth]{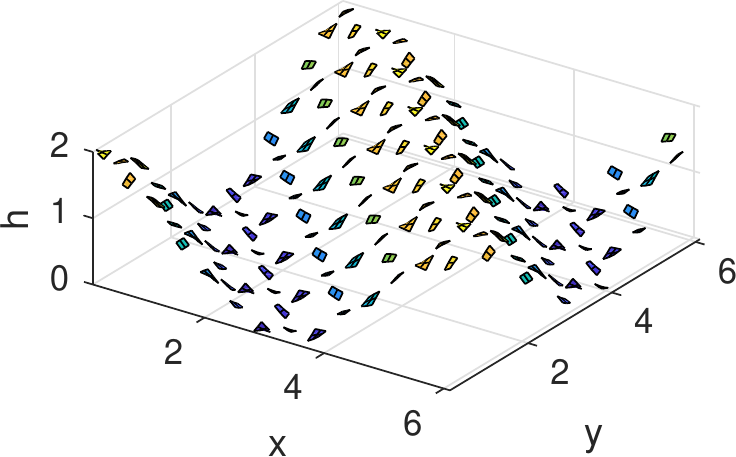}}
\end{tabular}
\end{figure}%
The micro-scale ideal wave \ode{}s~\cref{eqs:waveds} inside patches, with the linear interpolant patch coupling~\eqref{eqs:lint}, form a closed system.
\cref{fig:lint} shows some results of one simulation.
The simulation is on the \(2\pi\times 2\pi\) square domain, with \(14\times14\) patches.
Each patch has size ratio \(r=0.3\) (relatively large for visibility), and a \(7\times 7\) staggered microgrid (microgrid spacing \(d=0.045\)).
The initial condition plotted top-left of \cref{fig:lint} is that on the microgrid \(h=1+\sin(x+y)\), \(u=v=\frac1{\sqrt2}\sin(x+y)\).
Then \textsc{Matlab}'s \verb|ode23| integrates the system in time to the specified time (here about three-quarters of a period), giving the \(h\)-fields shown in \cref{fig:lint}.
Evidently the macro-scale wave propagates reasonably using this scheme, as confirmed by the analysis of \cref{sec:ecmclc}.

Observe in \cref{fig:lint}(b)--(c) that there appears to be some sub-patch structures superimposed upon the macro-scale \(\sin(x+y-\omega t)\) wave.
These sub-patch structures are fast waves on the micro-scale.
The reason these fast sub-patch waves occur is that the initial state of \cref{fig:lint}(a) is near but not quite on the slow subspace of the numerical patch scheme.
The displacement off the slow manifold then appears as the fast wave components we discern in the figure.
Such sub-patch fast waves may be removed by either appropriate initialisation of the macro-scale \cite[as discussed in geophysical fluid dynamics by, e.g.,][]{Leith80, Vautard86}, or by some physical micro-scale dissipation (\cref{sec:ildamv}).

\begin{figure}
\centering
\caption{\label{fig:idealevals}spectrum of eigenvalues for the staggered patch scheme with inter-patch coupling by (a)~linear interpolation, and (b)~spectral interpolation.  We plot all eigenvalues of the Jacobian of the system on a \(2\pi\times2\pi\) domain, with \(10\times10\) patches, each patch with \(7\times 7\)~microgrid, and of size ratio \(r=0.1\)\,.
The axes are quasi-log transformed: vertically \(\asinh(\Im\lambda)\); and horizontally \(\asinh(\Re\lambda\cdot10^7)\).}
\begin{tabular}{@{}cc@{}}
(a) nearest-neighbour linear
&(b) global spectral
\\\includegraphics[scale=0.9]{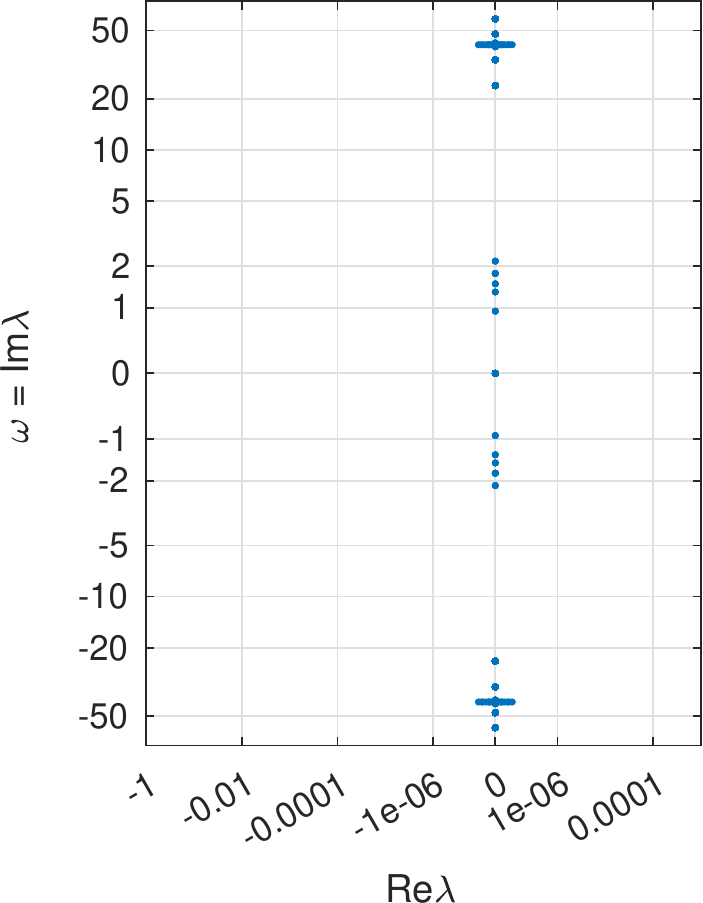}
&\includegraphics[scale=0.9]{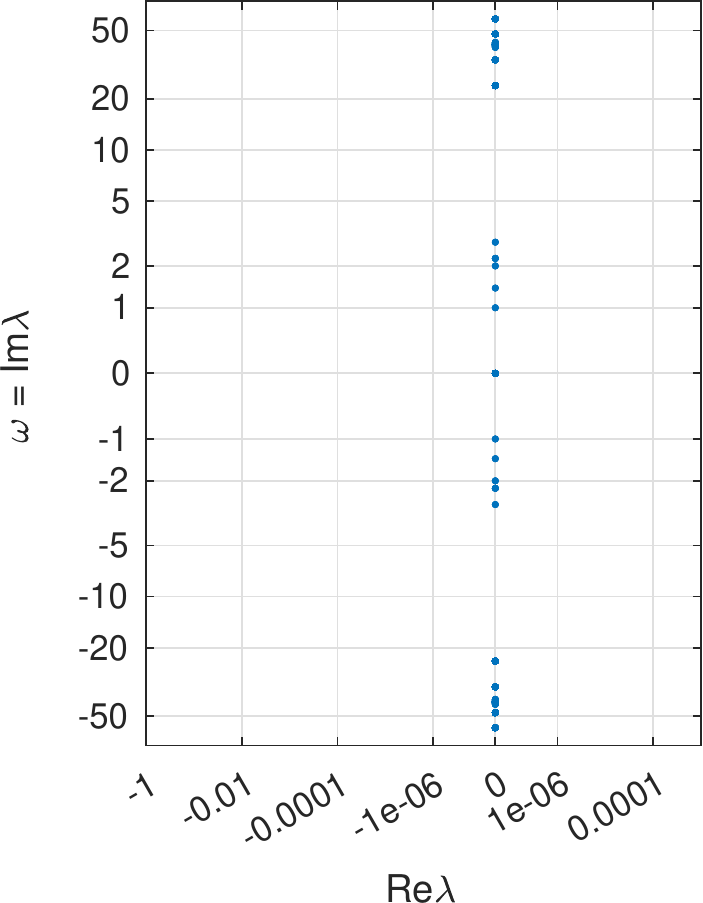}
\end{tabular}
\end{figure}%
We characterise the modes of the coupled-patch wave system by computing the eigenvalues of the Jacobian of the system.
Being linear, we construct the Jacobian of the system~\cref{eqs:waveds}+\eqref{eqs:lint} column-by-column by computing the time derivatives for a complete set of unit basis vectors for the state space of the system.
\cref{fig:idealevals}(a) displays the resultant eigenvalues in the complex plane (\(\asinh\) deformed) for a case with patch size-ratio \(r=0.1\).
The eigenvalues are all pure imaginary showing the dynamics of the coupled-patch system is entirely wave-like (there are a few modes with real-part of magnitude up to~\(10^{-7}\) which we interpret as due to round-off error in computing some of the multiply repeated eigenvalues).
The eigenvalues, or frequencies \(\omega=\Im\lambda\), fall into two broad categories separated by a spectral gap (see schematic \cref{fig:scalesep}(c)): slow modes with frequencies \(|\omega|\lesssim 3\); and fast waves with high frequencies \(|\omega|\gtrsim 20=\Ord{1/r}\).
The high frequencies characterise short wavelength sub-patch waves that are of no interest to the macro-scale dynamics, albeit, for example, visibly superimposed on the simulation of \cref{fig:lint}.
\footnote{An issue yet to be researched is whether the propagation of sub-patch fast waves across the domain, from patch to patch via the coupling, has any claim to some physically relevant meaning.  To date we have assumed not.}
This does not say that the short wavelength waves are negligible, just that  in linear systems they do not affect the macro-scale waves.  
For example, the \(5\)--\(20\)~second period waves that we enjoy at the beach barely affect the large scale tides of period \(12\)~hours (the short waves  only affect the relatively slow tides through nonlinear processes, see the discussion near the end of \cref{sslrnw}).

The slow modes, in turn, form two groups: the neutral modes with zero frequency; and the macro-scale wave modes with frequencies \(1 \lesssim |\omega| \lesssim 3\).
The case shown in \cref{fig:idealevals}(a) has \(1000\)~fast waves, \(427\)~neutral `vortical' modes, and \(48\)~macro-scale waves of interest.
With a \(10\times10\) staggered grid of patches there are \(5\times5\) macro-scale cells, and hence the macro-scale resolves \(2(5^2-1)=48\) wave modes. 
Among the neutral modes, three correspond to mean depth/flow, that is, each of \(h,u,v\) being independently constant across the domain.
The remaining neutral modes have zero~\(h\), 
\footnote{In vortical modes \(h\)-perturbations must be zero as otherwise gravity would act to push water around and hence be part of a wave of some finite frequency.}
and non-zero~\(u,v\) and represent vortical flow: for the case of \cref{fig:idealevals}(a), \(24\)~modes are macro-scale vortical flows, and \(400\)~modes are micro-scale sub-patch vortical flow.
This satisfactory structure in the eigenvalues\slash frequencies follows from imposing the patch scheme on the spatial wave dynamics of the discretised wave system~\eqref{eqs:waveds}.

\cref{fig:idealevals}(b) shows the same qualitative pattern of eigenvalues arise in the case of global spectral interpolation that we now describe.

\subsubsection{Global spectral interpolation}
\label{ssgsi}

As an alternative to the nearest-neighbour linear coupling, here we discuss the highly accurate spectral coupling of patches.
For a rectangular domain \(L_x\times L_y\), the procedure is to take the discrete Fourier transform of each 2D array of macrogrid values and use the standard Fourier-shift property to interpolate and provide patch-edge values.
For example, let's consider the \(h\)-field sampled as the macro-scale data \(H_{I,J}\) for odd~\(I,J\) on the macrogrid of spacing~\(D\) so that the \(H\)-sampling is of spacing~\(2D\).
Then the Fourier transform writes these as
\begin{equation}
H_{I,J}=\sum_{k,\ell}\tilde H_{k,\ell}e^{\i(kX_I+\ell Y_J)},
\label{eq:hft}
\end{equation}
for some \textsc{fft}-computed coefficients~\(\tilde H_{k,\ell }\), and for the appropriate set of wavenumbers (\(k,l\)~integer multiples of~\(2\pi/L_x,2\pi/L_y\), and \(|k|,|l|<\pi/(2D)\) where to avoid ambiguity we choose domain sizes~\(L_x,L_y\) to be odd multiples of~\(2D\)).

From the coefficients computed in~\eqref{eq:hft}, the macro-scale interpolated field is \(H(x,y)=\sum_{k,\ell }\tilde H_{k,\ell }e^{\i(kx+\ell y)}\).
Consequently for points relative to an \((I,J)\)th \(H\)-patch, say at \((X_I+\xi,Y_J+\eta)\) where \((\xi,\eta)\) are displacements from the centre of the \((I,J)\)th~patch to the edges of neighbouring~\(U\) and~\(V\) patches, the spectral interpolation gives micro-scale patch-edge values
\begin{equation}
h(X_I+\xi,Y_J+\eta):=\sum_{k,\ell }\big[\tilde H_{k,\ell }e^{\i(k\xi+\ell\eta)}\big]e^{\i(kX_I+\ell Y_J)}
\quad\text{for odd }I,J,
\label{eq:hift}
\end{equation}
computed over all patches via another \textsc{fft} for each~\((\xi,\eta)\).
Correspondingly interpolate the macro-scale \(U,V\)~fields to give patch-edge values for the micro-scale \(u,v\)~fields wherever needed.

Implementing such spectral inter-patch coupling, \cref{eq:hift} et al., and simulating the wave corresponding to that shown in \cref{fig:lint} we found the same qualitative macro-scale wave propagation.
An observable difference is that there are no fast waves superimposed on the simulation because, with spectral coupling, the initial condition that \(h,u,v\propto \sin(x+y)\) is precisely on the slow subspace of the multiscale staggered patch system.

As before, eigenvalues characterise the dynamics of the spectral patch scheme. 
We construct the Jacobian column-by-column from the time derivatives, as before, and then compute its eigenvalues as plotted in one case in \cref{fig:idealevals}(b).
The eigenvalues are all pure imaginary to within numerical round-off error, and hence all modes are either wave modes or neutral modes.
The macro-scale waves of interest are modes with frequencies, \(\omega=\Im\lambda\), in the range \(1\lesssim |\omega|\lesssim 3\).
Comparing these frequencies of the spectral-patch scheme with the frequencies of the micro-scale system~\cref{eqs:waveds}, they agree to errors less than \(2\cdot10^{-13}\).
That is, the spectral-patch scheme simulates macro-scale waves of the micro-scale system~\cref{eqs:waveds} to within numerical round-off error.
The patch scheme does this through coupling the micro-scale simulator that is computed only on small staggered patches in space.

\subsection{Eigenvalues confirm macro-scale consistency of linear coupling}
\label{sec:ecmclc}

The previous sections numerically explored the multiscale scheme when implemented with a finite number of patches on a finite spatial domain,~\(L_x\times L_y\), with a specified size ratio~\(r\).  
This section algebraically establishes for every~\(r\), and in an infinite spatial domain, that the long macro-scale waves over such coupled patches behave consistently and stably with the wave \pde{}s.
That is, simulations with this staggered patch scheme makes physically correct macro-scale predictions.

\begin{figure}
\centering
\begin{tabular}{@{}cc@{}}
\parbox[b]{15.5em}{\caption{\label{fig:onecell}the three staggered patches shown here form one cell that is doubly infinitely replicated in space, and coupled together.  The eigenvalues of the resultant system forms the basis for establishing general stability and consistency of macro-scale waves on the staggered patches.}}
&
\input{macroStaggerCell}
\end{tabular}
\end{figure}
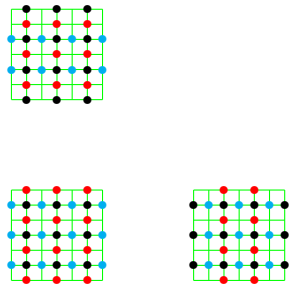

In the multiscale staggered grid of \cref{fig:patchMacro}, the basic cell is formed of the three patches illustrated in \cref{fig:onecell}.
Consider the scenario where this cell is doubly infinitely replicated across all 2D space.
Let the micro-scale wave dynamics on each patch be coupled together by the nearest-neighbour linear interpolation of \cref{ss:nnli}. 

To explore the dynamics of macro-scale, long, waves on these coupled patches we seek solutions\({}\propto\exp[\i(kI+\ell J)+\lambda t]\) where \((I,J)\in\ZZ^2\) are the indices of the staggered patches (\cref{fig:patchMacro}), where the frequency of any waves is \(\omega=\Im\lambda\), and where the wavenumber of large scale structures across the cells is~\((k,\ell)\). 
On the array of cells these wavenumbers are restricted to \(|k|,|\ell|\leq\pi/(2D)\).
For modes which have smooth sub-patch structures, the wavenumber~\((k,\ell)\) is the wavenumber of the predicted long macro-scale waves.
For modes with significant sub-patch structure, \((k,\ell)\)~is the wavenumber of large-scale \emph{modulations} to the fast micro-scale sub-patch waves.
We focus on the dynamics of the long macro-scale waves that have smooth sub-patch structure.

On this infinite domain the only macro-scale length scale is the separation,~\(D\), of the patches, so for the rest of this subsection we non-dimensionalise the problem so that in effect \(D=1\).

%

The Jacobian of the linear patch scheme for the discretised linear waves~\eqref{eqs:waveds} encodes all the dynamics of modes in \(\exp[\i(kI+\ell J)+\lambda t]\).
To construct an analytic Jacobian to analyse for the multiscale patch scheme, \cref{ss:nnli}, on an infinite array of patches and for general~\(r\), we recognise that the Jacobian is linear in patch size~ratio~\(r\) and in macro-scale variations~\(e^{\pm\i2k}\) and~\(e^{\pm\i2\ell}\).
By computing the Jacobian of the numerical time-derivative function, for various values of~\(r,k,\ell\), it was straightforward to both fit and confirm a correct algebraic expression for the Jacobian of each patch configuration for waves.  
The ancillary material, \cref{sec:mcptsw}.5, lists the Jacobian for the specific case of patches with \(7\times7\) micro-scale sub-patch grid (\(n=6\), and illustrated by \cref{fig:patchMacro,fig:onecell}).
In this specific case, per cell of \cref{fig:onecell}, the micro-scale discretisation~\eqref{eqs:waveds} applies at \(59\)~interior points of the three patches in a cell (\(21\)~for \(\dot h_{i,j}\), and \(19\)~each for \(\dot u_{i,j}\) and~\(\dot v_{i,j}\)).
Thus in this case the Jacobian is~\(59\times59\) for each~\(r,k,\ell\) (as listed in \cref{sec:mcptsw}.5).

We constructed and analysed Jacobians for the sub-patch microgrid cases \(n=2,6,10,14\) (Jacobians of size \(m\times m\) for \(m=3,59,187,387\) respectively).
The macro-scale wave results for all these cases appear to be straightforward variations of the results for the case \(n=6\)\,, so we discuss this specific case.

So set \(n=6\) and consider the \(59\times 59\) Jacobian of the dynamics in a cell with long wavelength, macro-scale, modulation.  
For every given~\(r,k,\ell\) numerical computation showed the \(59\)~eigenvalues are of three types: \(40\)~sub-patch fast waves with large eigenvalues; \(16\)~zero eigenvalues of vortices; and \(3\)~small eigenvalues of the slow macro-scale modes that have smooth sub-patch structure.

We derive an analytic asymptotic approximation to the macro-scale modes of the patch scheme by seeking the small eigenvalues as an expansion in small wavenumbers~\(k,\ell\).
Unfortunately, the uninteresting \(16\)~vortical modes are in the same  eigenspace as the three macro-scale modes.
So to construct a basis for the 19D slow eigenspace (a basis which requires generalised eigenvectors) we have to ensure that the basis decouples the \(16\)~vortical modes from the three macro-scale modes.
For the \(n=6\) case of the Jacobian~\cJ\ we seek to satisfy the eigenspace equation \(\cJ\cV=\cV\cL\) with basis vectors in the \(19\)~columns of~\cV, and a \(19\times19\) matrix~\cL, such that the matrix~\cL\ has the partitioned form, in which \(x\)~denotes the only nonzero entries,
\begin{equation*}
\cL=\left[\begin{array}{ccc|ccc} 0&x&x&0&\cdots&0\\
 x&0&0&0&\cdots&0\\
 x&0&0&0&\cdots&0\\
 \hline
 x&x&x&0&\cdots&0\\
 \vdots&\vdots&\vdots&\vdots&&\vdots\\
 x&x&x&0&\cdots&0\\
 \end{array}\right].
\end{equation*}
Given such a form for~\cL\ we know that the first three columns of~\cV\ are of three modes---macro-scale waves modes---that evolve independently of the other \(16\)~modes---sub-patch vortical modes.
The computer algebra of \cref{sec:mcptsw}.6 constructs an asymptotic expansion for~\cV\ and~\cL\ valid for small wavenumbers~\(k,\ell\).
More precisely, the computer algebra code defines \(\cJ_0:=(2r/n)J|_{k=\ell=0}\) and \(\cJ_1:=(2r/n)\cJ-\cJ_0\) and then seeks properties of \(\cJ_0+\epsilon\cJ_1\) as a power series in the homotopy parameter~\(\epsilon\) to give approximations valid for small wavenumbers~\(k,\ell\), that is, for long macro-scale modulations and waves. 

Then the eigenvalues of the top-left \(3\times3\) block of~\cL\ determines the dynamics of the macro-scale waves uncoupled from the influence of all the \(40\)~sub-patch fast waves and the \(16\)~sub-patch vortical modes.
The computer algebra of \cref{sec:mcptsw}.6 determines the top-left \(3\times3\) block is
\begin{align*}
\cL_{3\times3}&=\begin{bmatrix} 0
& - \i\sin(k)e^{-\i k}
& - \i\sin(\ell)e^{-\i\ell}
\\
-\i\sin(k)e^{+\i k} &0 &0\\
-\i\sin(\ell)e^{+\i\ell} &0 &0 \end{bmatrix}
\\&\quad{}
\times\left\{\epsilon
+\epsilon^3\tfrac4{27}r^2(\sin^2k+\sin^2\ell)
+\epsilon^5\tfrac{16}{243}r^4(\sin^2k+\sin^2\ell)^2
+\Ord{\epsilon^6}
\right\} .
\end{align*}
Evidently the top-left block factors into a wave-like matrix multiplied by a scalar factor (within~\(\{\cdot\}\)).
The scalar factor shows that the power series in~\(\epsilon\) is essentially a power series in either the patch size ratio~\(r\) or small wavenumber~\((k,\ell)\): so set the homotopy parameter \(\epsilon=1\).
The above matrix factor has characteristic equation \(-\lambda(\lambda^2+\sin^2k+\sin^2\ell)=0\)\,.
Consequently, in the patch scheme the invariant subspace corresponding to~\(\cL_{3\times3}\) represents firstly, with eigenvalue zero, a macro-scale vortical mode, and secondly, via the pair of complex conjugate eigenvalues, a macro-scale wave with dispersion relation
\begin{align}
\omega^2&=(\sin^2k+\sin^2\ell)\left\{1
+\tfrac4{27}r^2(\sin^2k+\sin^2\ell)
+\tfrac{16}{243}r^4(\sin^2k+\sin^2\ell)^2
+\cdots
\right\}^2
\nonumber\\&
=(k^2+\ell^2)
-\tfrac13\left[k^4+\ell^4-\tfrac89r^2(k^2+\ell^2)^2\right]
+\cdots\,.
\label{eq:w1dr}
\end{align}
The above pattern is the same for the other values of~\(n\) explored.
\begin{itemize}
\item That the dispersion relation is \(\omega^2\approx k^2+\ell^2\) establishes that the macro-scale waves in the staggered patch scheme reasonably represent the long waves of the linear wave \pde~\eqref{eqs:wavepdes}---recall that long waves are those with \(|k|,|\ell|\leq\pi/(2D)\) (here we scaled \(D=1\)).
\item The quartic correction term~\(-\tfrac13(k^4+\ell^4)\) in the dispersion relation~\eqref{eq:w1dr} shows that the macro-scale waves in the patch scheme have a small anisotropy due to the square grid of patches \emph{and} their coupling via linear interpolation.
\item The \(r^2\) quartic correction term~\(+\tfrac8{27}r^2(k^2+\ell^2)^2\) reflects that the underlying micro-scale lattice system~\eqref{eqs:waveds} approximates the wave \pde~\eqref{eqs:wavepdes}---as patches get smaller, \(r\to0\), the lattice system\({}\to{}\)the \pde\ and this correction goes to zero.
Similarly for other \(r\)-effects in the dispersion relation~\eqref{eq:w1dr}.
\end{itemize}
Thus the dispersion relation~\eqref{eq:w1dr} of the macro-scale waves in the staggered patch scheme establishes that the linear coupling scheme is consistent with the underlying wave systems, both discrete~\eqref{eqs:waveds} and continuous~\eqref{eqs:wavepdes}.

\cref{ssgsi} also introduces coupling the staggered patches with a global spectral interpolation.
Numerically we found that the macro-scale waves in the spectral scheme reproduced the large scale waves in the lattice system~\eqref{eqs:waveds} to numerical round-off error: that is, the only observable error is in the micro-scale discretisation of the wave \pde~\eqref{eqs:wavepdes}, and none at all in the patch scheme.
The reason for this numerically zero error is that every macro-scale wave, \({}\propto\exp[\i(kI+\ell J)]\), is represented exactly in a spectral interpolation, and so all the edge values of the patches in the cell of \cref{fig:onecell} are exact for the macro-scale wave.
Hence, by the homogeneity of the underlying lattice system~\eqref{eqs:waveds}, the corresponding sub-patch structures also exactly match~\(\exp[\i(kI+\ell J)]\) and so give the numerically exact dispersion relation for the macro-scale waves.

Ongoing research aims to find a variety of coupling schemes for the staggered patch scheme, other than the two explored here, which have good macro-scale accuracy, and are also robustly stable.

\section{Include small drag and micro-scale viscosity}
\label{sec:ildamv}

The previous \cref{sec:spsm2d} explores the good behaviour of a staggered patch scheme for the ideal wave \pde~\cref{eqs:wavepdes}, and its micro-scale discretisation~\cref{eqs:waveds}.
This section explores the patch scheme for such a wave system with small micro-scale dissipation included: here we include both a drag and a viscosity---both linear.
The aim is to see, before grappling with nonlinear problems, what issues we can resolve for the physically interesting scenario where the micro-scale has some dissipation which is often negligible on the macro-scale of interest. 
This section further develops two extremes: firstly, what is the `simplest' patch scheme that nonetheless retains qualitative accuracy for the waves; and secondly, a highly accurate spectral interpolation patch scheme.

This section considers the wave \pde\ with dissipation
\begin{subequations}\label{eqs:wavepded}%
\begin{align}
&\D th=-\D xu-\D yv\,,
\\&\D tu=-\D xh -c_Du+c_V\delsq u\,,
\\&\D tv=-\D yh -c_Dv+c_V\delsq v\,,
\end{align}
\end{subequations}
where \(c_D\)~is the coefficient of drag on the `flow', and \(c_V\)~is the coefficient of `fluid viscosity'.
On the staggered microgrid~\eqref{eqs:waveds}, with micro-scale spacing~\(d\), we generally discretise these \pde{}s by the straightforward centred difference approximations
\begin{subequations}\label{eqs:wavedd}%
\begin{align}
\bullet&
\de t{h_{i,j}}=-\frac{u_{i+1,j}-u_{i-1,j}}{2d}-\frac{v_{i,j+1}-v_{i,j-1}}{2d}\,,
\\\color{cyan}\bullet&
\de t{u_{i,j}}=-\frac{h_{i+1,j}-h_{i-1,j}}{2d} -c_Du_{i,j}
+c_V\frac{u_{i+2,j}+u_{i-2,j}+u_{i,j+2}+u_{i,j-2}-4u_{i,j}}{4d^2}\,,
\\\color{red}\bullet&
\de t{v_{i,j}}=-\frac{h_{i,j+1}-h_{i,j-1}}{2d} -c_Dv_{i,j}
+c_V\frac{v_{i+2,j}+v_{i-2,j}+v_{i,j+2}+v_{i,j-2}-4v_{i,j}}{4d^2}\,.
\end{align}
\end{subequations}

\begin{figure}
\centering
\caption{\label{fig:dampevals}eigenvalue spectrum for the staggered patch schemes for drag \(c_D=10^{-6}\) and viscosity \(c_V=10^{-4}\).  We plot, on quasi-log axes, all eigenvalues of the Jacobian of the system on a \(2\pi\times2\pi\) domain, with \(10\times10\) patches, each of size ratio \(r=0.1\) with an \(13\times13\) microgrid.  The two plotted spectra are for two different macro-scale interpolations.}
\begin{tabular}{@{}cc@{}}
(a) local low-order
&(b) global spectral
\\ \includegraphics[scale=0.9]{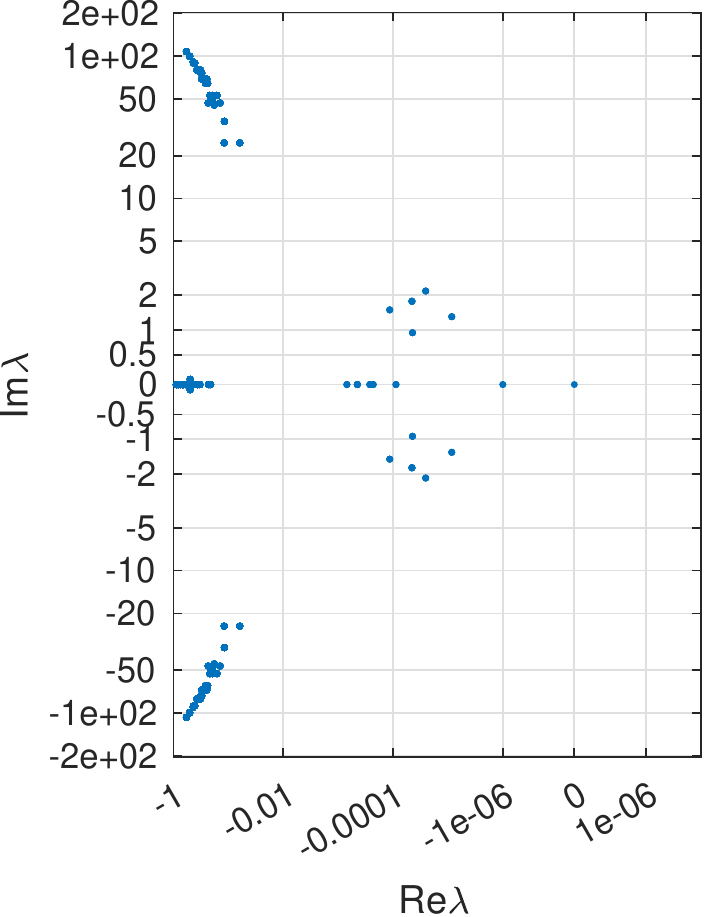}
& \includegraphics[scale=0.9]{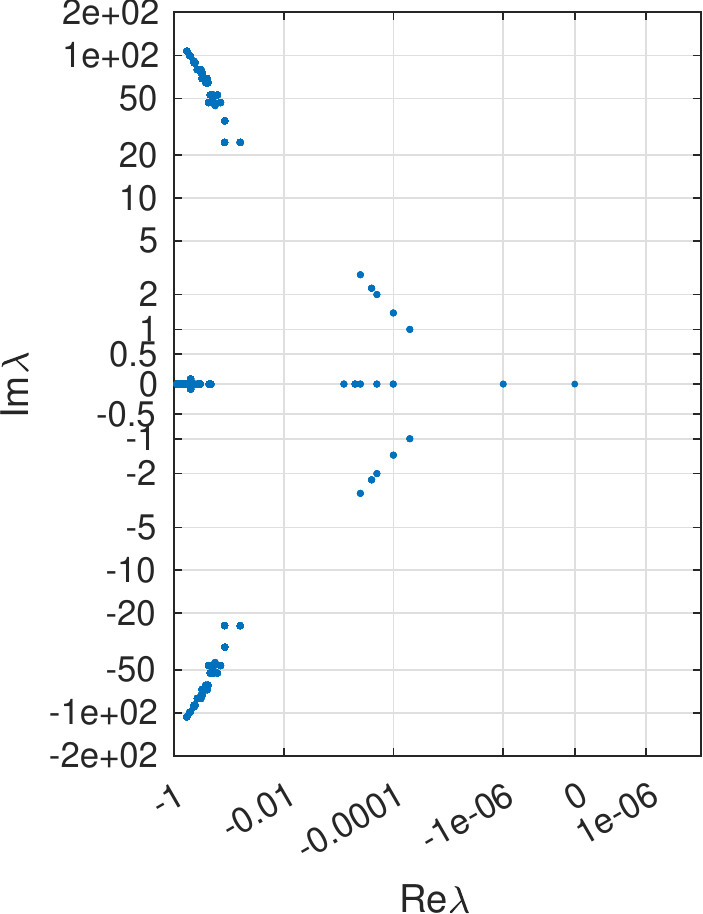}
\end{tabular}
\end{figure}

The discretisation~\cref{eqs:wavedd} is qualified by ``generally'' because in a patch scheme and for microgrid points near to the edge of a patch, sometimes there is no natural value for the required field at indices \((i\pm2,j)\) or \((i,j\pm 2)\).
Where this deficiency occurs in the patches shown in \cref{fig:patchMacro,fig:onecell} we have to code an alternative for the viscous dissipation.
In regard to coding the micro-scale viscous dissipation on microgrid lines neighbouring the patch-edges we explored various alternatives, including: setting \(\delsq=0\) when it is not available;  setting to zero unknown first differences in the discrete formula for~\(\delsq\); setting to zero unknown second differences in the discrete formula for~\(\delsq\) (equivalent to local linear extrapolation); and assuming the micro-scale field values `wrap around' to the opposite patch edge (helps damp sub-patch shear).
Many patch and interpolation designs were found to be unstable because such \(\delsq\)-coding, in the multiscale geometry of the patch scheme, causes some eigenvalue real-parts to become positive.
Plots of their spectrum (like \cref{fig:idealevals,fig:dampevals}) indicated that none of these alternatives were completely satisfactory, although the last alternative was mostly good.
Ongoing research is exploring a range of other alternatives.

\begin{figure}
\centering\caption{\label{fig:patchMacroX}2D staggered macro-scale grid of patches for the weakly dissipated wave-like system~\cref{eqs:wavepded}.
The patch centre-value determines its nature:
\({\color{black}\bullet}~H\)-patch;
\({\color{cyan}\bullet}~U\)-patch; and
\({\color{red}\bullet}~V\)-patch.
In comparison with \cref{fig:patchMacro}, the difference is that here the patch-edge is \emph{two} microgrid points thick: that is, we interpolate macro-scale values to a perimeter \emph{two} microgrid points thick around each patch, and here solve  \pde{}s on the \(5\times5\) interior of a \(9\times9\) patch.}
\input{macroStaggerPatchesX}
\end{figure}
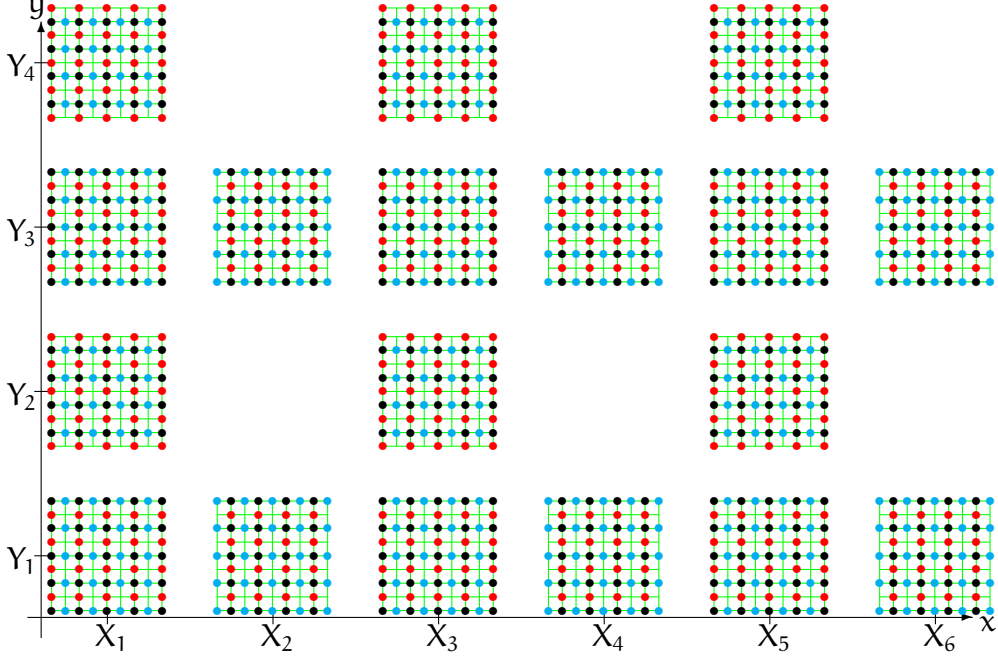

The most straightforward alternative that is so far successful is the following.
\cref{fig:patchMacroX} indicates the change: we take the grid of \cref{fig:patchMacro} and add another microgrid perimeter around each patch so that now the \emph{edge} of a patch is \emph{two} microgrid points thick.
The scheme is then to interpolate macro-scale values, the patch-centre values, to provide edge values in the two-point thick edge around each patch; and to evaluate the \pde\ discretisations~\eqref{eqs:wavedd} in the interior.  
For the example patches of \cref{fig:patchMacroX} that each have a \(9\times9\) microgrid, the \pde\ discretisation~\eqref{eqs:wavedd} would be evaluated in the \(5\times5\) interior of each patch, and macro-scale interpolation gives the \(2\times9\) strips of patch-edge values.

\cref{fig:dampevals} illustrates the success of this patch coupling scheme for weakly dissipating waves with drag \(c_D=10^{-6}\) and viscosity \(c_V=10^{-4}\).
The figure plots the spectrum of the scheme when used on a \(10\times10\) array of square patches in space.
Each patch is of size ratio \(r=0.1\) so the \pde\ discretisation~\eqref{eqs:wavedd} is computed on only a fraction \(3r^2=3\%\) of the domain.
Each patch is formed with a \(13\times 13\) microgrid: a \(9\times9\) interior in which~\eqref{eqs:wavedd} is computed; and four strips \(2\times13\) around the edge where interpolation couples the patches.
Then \cref{fig:dampevals} plots the resultant \(4675\) eigenvalues forming the spectrum of the system.
No eigenvalue has positive real-part so the scheme is stable.
The eigenvalues occur in seven clusters.
\begin{itemize}
\item There are three superslow eigenvalues of `mean flow': one~\(0\) reflects conservation of~\(h\); two at~\(-c_D=-10^{-6}\) reflect drag on mean velocity.

\item There are three clusters of `slow' macro-scale eigenvalues in the range \(-10^{-3}=-Nc_V<\Re\lambda<-\tfrac12c_V=-0.5\cdot10^{-4}\).
The cluster with \(\Im\lambda=0\) represent \(24\)~macro-scale vortical flow modes in \(u,v\), slowly dissipating, with no \(h\)-component.
The two clusters with non-zero~\(\Im\lambda\) are the \(24\)~pairs of complex conjugate eigenvalues of the slowly dissipating macro-scale waves (recall that on a \(10\times10\) staggered grid of patches there are \(5\times5\) macro-scale cells, and hence the macro-scale resolves \(5^2-1=24\) waves-pairs).  
These are the modes of main physical interest.

\item There is a useful spectral gap to the next three clusters of `fast' eigenvalues, \(4600\)~of them, representing micro-scale, sub-patch, dynamics.  
The two clusters with \(|\Im\lambda|>20\) are the \(1600\)~pairs of complex conjugate eigenvalues of the rapidly dissipating micro-scale sub-patch fast waves.  
The cluster with \(\Im\lambda\approx0\) represent \(1400\)~micro-scale sub-patch vortical flow modes.
The modes in this cluster decay at a relatively rapid rate~\(\Ord{c_V\pi^2/r^2}=\Ord{0.1}\).
All these fast modes are of little interest as they dominantly reflect our artifice of imposing a multiscale patch structure on the wave problem.
\end{itemize}
Thus this staggered patch scheme's eigenvalues form the physically appealing spectrum shown in \cref{fig:dampevals}.

The accuracy of the macro-scale waves in the local low-order coupling appears poor (\cref{fig:dampevals}(a)), and may repay further investigation. 
The virtue of the local low-order coupling is that it retains the correct structure of the multiscale wave dynamics in a coupling scheme that is the simplest to implement. 

The eigenvalues plotted in \cref{fig:dampevals}(b) indicate that the weakly damped macro-scale waves arising with spectral coupling are highly accurate.
For waves proportional to \(\exp[\lambda t+\i(kx+\ell y)]\), the characteristic equation of the wave \pde~\eqref{eqs:wavepded} is
\begin{equation}
\big[\lambda+c_D+c_V(k^2+\ell ^2)\big]
\left\{ \lambda^2 +\big[c_D+c_V(k^2+\ell ^2)\big] +k^2+\ell ^2 \right\}=0\,.
\label{eq:chareqn}
\end{equation}
Over several computational experiments, for each macro-scale wave eigenvalue~\(\lambda_j\) we determine the integer \(k^2+\ell ^2\) by rounding~\(\lambda_j^2\), and then find that the characteristic polynomial
\begin{equation*}
\lambda^2 +\big[c_D+c_V(k^2+\ell ^2)\big] +(k^2+\ell ^2) = \Ord{d^2}
\quad\text{as }r\to0\,.
\end{equation*}
That is, evidently the spectrally coupled patch scheme for the wave \pde~\eqref{eqs:wavepded} has the correct macro-scale wave eigenvalues to the error inherent in the coded micro-scale discretisation~\eqref{eqs:wavedd}.

\section{Nonlinear turbulent flood 2D simulation}
\label{sec:ntfs}

Direct numerical simulation of the complexities of turbulent floods over any reasonable physical domain of interest is far too detailed to be yet feasible.
However, we have previously derived shallow water models based upon the Smagorinski \cite[]{Cao2014b}, and the \(k\)-\(\epsilon\) turbulence models \cite[]{Roberts99c}.
So in this section we take a step towards direct numerical simulation in patches by applying the patch scheme to the nonlinear, Smagorinski-based, shallow water model.


\subsection{Model turbulent floods via a Smagorinski closure}
\label{sec:mtfsc}

Starting with the Smagorinski turbulence closure for 3D turbulent fluid flow \cite[e.g.,][]{Ozgokmen2007a}, \cite{Cao2014b} used centre manifold theory \cite[e.g.,][]{Roberts1988, Potzsche:2006uq} to justify and construct a `shallow water' model in terms of depth averaged velocities: we emphasise that these are not ``depth averaged equations'' but are  the result of a systematic centre manifold modelling which is written in terms of ``depth averaged quantities''. 
In terms of the depth~$h(x,y,t)$, and the depth-averaged velocities~$\uu(x,y,t)$ and~$\vv(x,y,t)$ (with mean flow speed $\bq=\sqrt{\uu^2+\vv^2}$) the derived \pde{}s are 
\begin{subequations}\label{eqs:Intro}
\begin{align}
\D{t}{h}\approx&{} -\D{x}{}\left(h\uu\right)-\D{y}{}\left(h\vv\right)
,\label{sed:IntroH}
\\
\D{t}{\uu}\approx&{} -0.00293\frac{\bq\uu}{h} -0.993\D{x}h
-1.045\uu\D x\uu-1.017\vv\D y\uu
+0.094\bq h\delsq\uu
\,,\label{sed:IntroU}
\\
\D{t}{\vv}\approx&{} -0.00293\frac{\bq\vv}{h} -0.993\D{y}h
-1.045\vv\D y\vv-1.017\uu\D x\vv 
+0.094\bq h\delsq\vv
\,.\label{sed:IntroV}
\end{align}
\end{subequations}
These nonlinear \pde{}s encode the principal physical processes in large scale floods and tsunamis.
\pde~\eqref{sed:IntroH} represents conservation of water.
\pde~\eqref{sed:IntroU} governs momentum in the horizontal \(x\)-direction:
\(\bq\uu/h\)~is a turbulent bed-drag; \(\D xh\)~the out-of-equilibrium driving by hydrostatic pressure; \(\uu\D x\uu\) and~\(\vv\D y\uu\) is the effective advection of the velocity profile (non-constant in the vertical); and \(\bq h\delsq\uu\) is the effective horizontal mixing that is a combination of direct turbulent mixing and a Troutan-like effect \cite[p.143, e.g.]{Ribe01}.
Similarly for \pde~\eqref{sed:IntroV} and momentum in the horizontal \(y\)-direction.

The detailed mathematical derivation of the \pde{}s~\eqref{eqs:Intro} generated more terms in its systematic asymptotic expansions \cite[]{Cao2014b}.
However, neglecting terms with small coefficients that appear to have negligible effect on predictions, we arrive at the \pde\ system~\eqref{eqs:Intro}.

Our coupled staggered patch scheme for the nonlinear wave \pde\ system~\eqref{eqs:Intro} is the following.
First, on a micro-scale lattice in 2D space, code a micro-scale staggered discretisation of the \pde{}s~\eqref{eqs:Intro}.
Second, code these micro-scale discretisations into the staggered patches of the scheme of \cref{fig:patchMacroX}: the figures shown for this section are all for the particular case of a \(9\times9\) microgrid in each patch, and the \pde{}s~\eqref{eqs:Intro} discretised on the \(5\times5\) interior (as in \cref{fig:patchMacroX}), but for patch-size ratio \(r=0.1\) (much smaller than in the figure).
Third, couple the patches with spectral interpolation from the macro-scale-lattice of patch-centre values to the edge-values of each patch.
This then gives a function that computes time derivatives of the \(5\times5\) dynamic variables in each and every patch, coupled together.

\begin{figure}
\centering
\caption{\label{fig:turbevals}eigenvalue spectrum for the staggered patch scheme for the turbulent shallow water model~\eqref{eqs:Intro}.  We plot, on quasi-log axes, all eigenvalues of the Jacobian of the linearised system on a \(2\pi\times2\pi\) domain, with \(10\times10\) patches, each patch of size ratio \(r=0.1\) with an \(9\times9\) microgrid.  The two spectra are for the linearisation about a depth \(h=1\) and the listed mean~flow.}
\begin{tabular}{@{}cc@{}}
(a) \(\uu=\vv=10^{-4}\)
&(b) \(\uu=\vv=0.1\)
\\\includegraphics[scale=0.9]{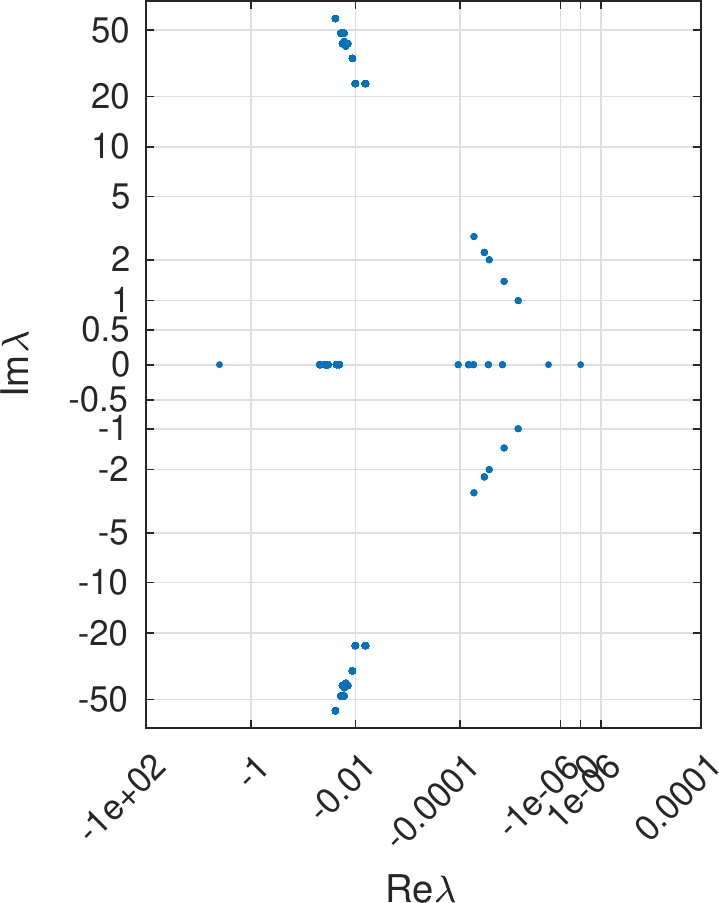}
& \includegraphics[scale=0.9]{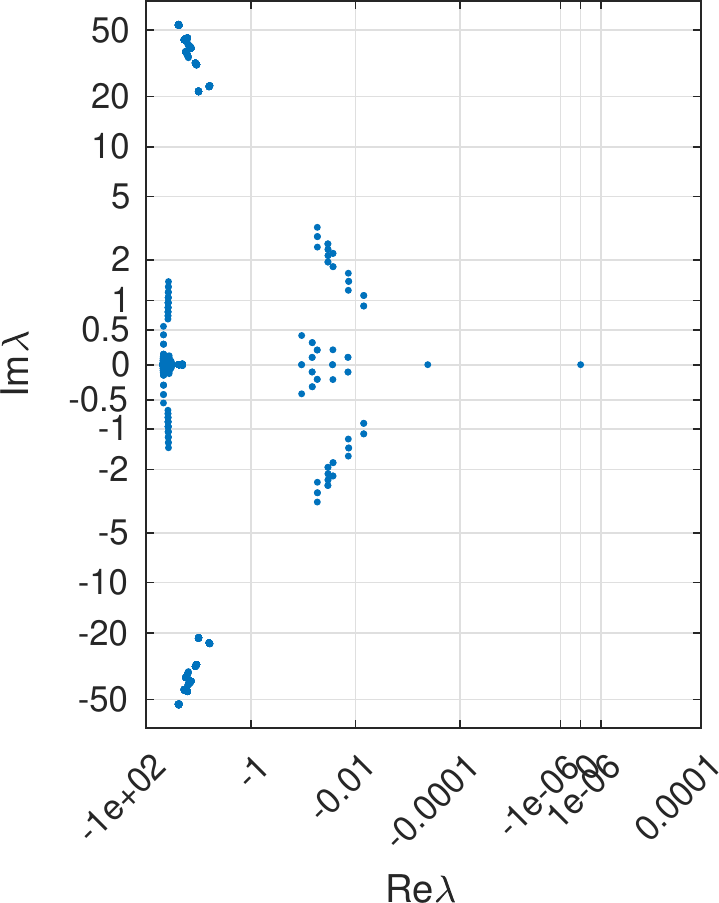}
\end{tabular}
\end{figure}%
As a first exploration of the patch scheme we discuss the linearised dynamics about the quasi-equilibrium of fixed water depth, non-dimensionally \(h=1\), and mean flow of constant~\((\uu,\vv)\).
The Jacobian of the patch scheme is formed by numerical differentiation (with step~\(10^{-8}\)) of the coded time derivative function, and then the eigenvalues of the Jacobian are found.
\cref{fig:turbevals} plots two cases, both with a \(10\times10\) macro-scale grid of patches.
\begin{itemize}
\item \cref{fig:turbevals}(a) plots the spectrum of eigenvalues for small underlying mean flow, and shows that the turbulent dissipation is much like the case of \cref{sec:ildamv} where ideal waves have weak superimposed drag and viscous damping (\cref{fig:dampevals}(b)).

\item \cref{fig:turbevals}(b) plots the spectrum of eigenvalues for medium mean flow, \(\uu=\vv=0.1\), and shows a broadly similar overall structure of seven clusters, but dissipating more rapidly due to the higher level of turbulent mixing in this mean flow.  
However, the detail is perturbed by the mean flow. 
In particular, the imaginary parts of the macro-scale eigenvalues are changed due to the advection of the underlying waves and vortices by the underlying mean flow.
\end{itemize}
The spectra for other cases and parameter values are similarly satisfactory.

\subsection{Linearisation is relevant for nonlinear waves}
\label{sslrnw}

\cref{fig:turbevals} shows spectral gaps between the eigenvalues of the macro-scale waves and the micro-scale sub-patch structures: 
in~(a) the gap is \(-0.01\lesssim\Re\lambda\lesssim-0.0001\); 
in~(b) the gap is \(-10\lesssim\Re\lambda\lesssim-0.1\).
Let's consider flow regimes where such spectral gaps occur in the linearised dynamics about a point of physical interest.
By centre manifold theory \cite[e.g.,][]{Haragus2011, Roberts2018a}, with some adaptations, we are then assured that around each such point there exists a domain of state space in which important properties hold for the nonlinear waves on the coupled staggered patches. 
Firstly, in the domain there exists a centre manifold of the out-of-equilibrium nonlinear macro-scale wave dynamics corresponding to the cluster of eigenvalues with small real-part \cite[e.g.,][\S2.3.1]{Haragus2011}.
Secondly, this nonlinear centre manifold is exponentially quickly attractive through the decay of the micro-scale sub-patch structures corresponding to the eigenvalues of large negative real-part \cite[e.g.,][\S2.3.4]{Haragus2011}.
Thirdly, the domain is of a finite size which may be bounded from below \cite[Lem.~12]{Roberts2018a}.
That is, the dynamics of the coupled patch scheme typically is attracted exponentially quickly, through the damping of micro-scale sub-patch waves, to the dynamics of macro-scale waves.

\paragraph{The union of local theory forms a global theory}
This emergence of macro-scale wave dynamics in the staggered patch scheme applies in a domain around each and every point of state space for which the linearised spectrum is like \cref{fig:turbevals}.
The union of these domains forms a global domain in which the union of the local centre manifolds form a global and attractive centre manifold.  
That is, over any parameter regime where the spectra are like \cref{fig:turbevals}, we are assured that macro-scale waves generally emerge in the multiscale patch scheme simulations.

\paragraph{When micro-scale dissipation is negligible}
However, theoretical support is much more delicate in regimes where the micro-scale dissipation is negligible \cite[e.g.]{Lorenz92}.
In regimes where all eigenvalues have effectively zero real-part, where some eigenvalues are large and some are small, representing fast waves and slow waves respectively, such as \cref{fig:idealevals}, then we are reasonably assured that there exist systems arbitrarily close to a given nonlinear system, here the patch scheme, which possess a slow manifold \cite[\S2.5]{Roberts2018a}.
The slow manifold corresponds to the set of small eigenvalues, such as those for  \(|\lambda|<3\) in \cref{fig:idealevals}, and so here the slow manifold consists of the macro-scale waves, albeit also with sub-patch vortices.
One caveat is that such a slow manifold is not attractive. 
For nonlinear systems in general the evolution of the slow modes, the macro-scale waves, is affected by the presence of any fast sub-patch waves---an effect which is typically quadratic in the fast wave amplitude \cite[Ch.~13, e.g.]{Roberts2014a}.
Thus in regimes where sub-patch dissipation is negligible, one needs to ensure that sub-patch waves are small enough not to significantly affect the macro-scale, through the nonlinearities, on the time-scale of interest.

\paragraph{Consistency}
That the macro-scale waves simulated in the staggered patch scheme appropriately predict the macro-scale dynamics of the underlying micro-scale system follows from the consistency of their eigenvalues with the eigenvalues of the \pde{}s that was established by \cref{sec:spsm2d,sec:ildamv}.

\subsection{Simulate turbulent floods in 2D space}

After the development and testing summarised previously, the resulting code (listed in \cref{sec:mcptsw}) is here used in some example simulations.
All these particular simulations use global spectral interpolation to couple the staggered patches in a \(2\pi\times2\pi\), doubly periodic, spatial domain.

\begin{figure}
\centering\caption{\label{fig:sim1t0}initial condition of a small-amplitude progressive wave across the domain of crest-trough height~\(0.1\), with patches of size ratio \(r=0.4\) so we can see the patches: bottom-left is the water depth; top-left and bottom-right are the~\(v\) and~\(u\) velocities, respectively (scaled by~\(20\) for clarity).   
\cref{fig:sim1t8} plots the predicted wave at time \(t=8\).}
\includegraphics[scale=0.9]{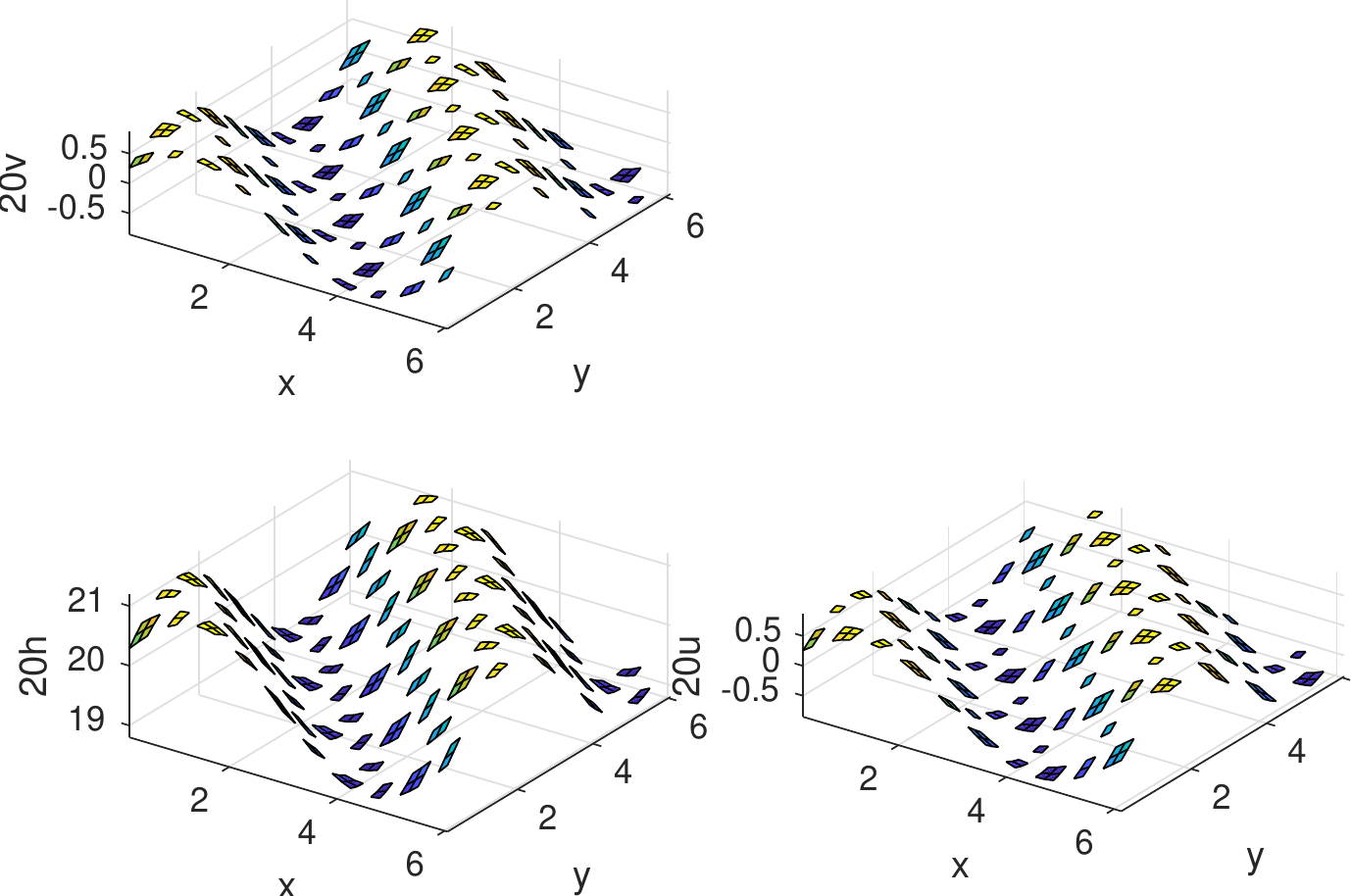}
\end{figure}%
\begin{figure}
\centering\caption{\label{fig:sim1t8}from the initial condition of \cref{fig:sim1t0}, approximately two periods later, at time \(t=8\), the progressing wave shows the development of some nonlinear distortion.}
\includegraphics[scale=0.9]{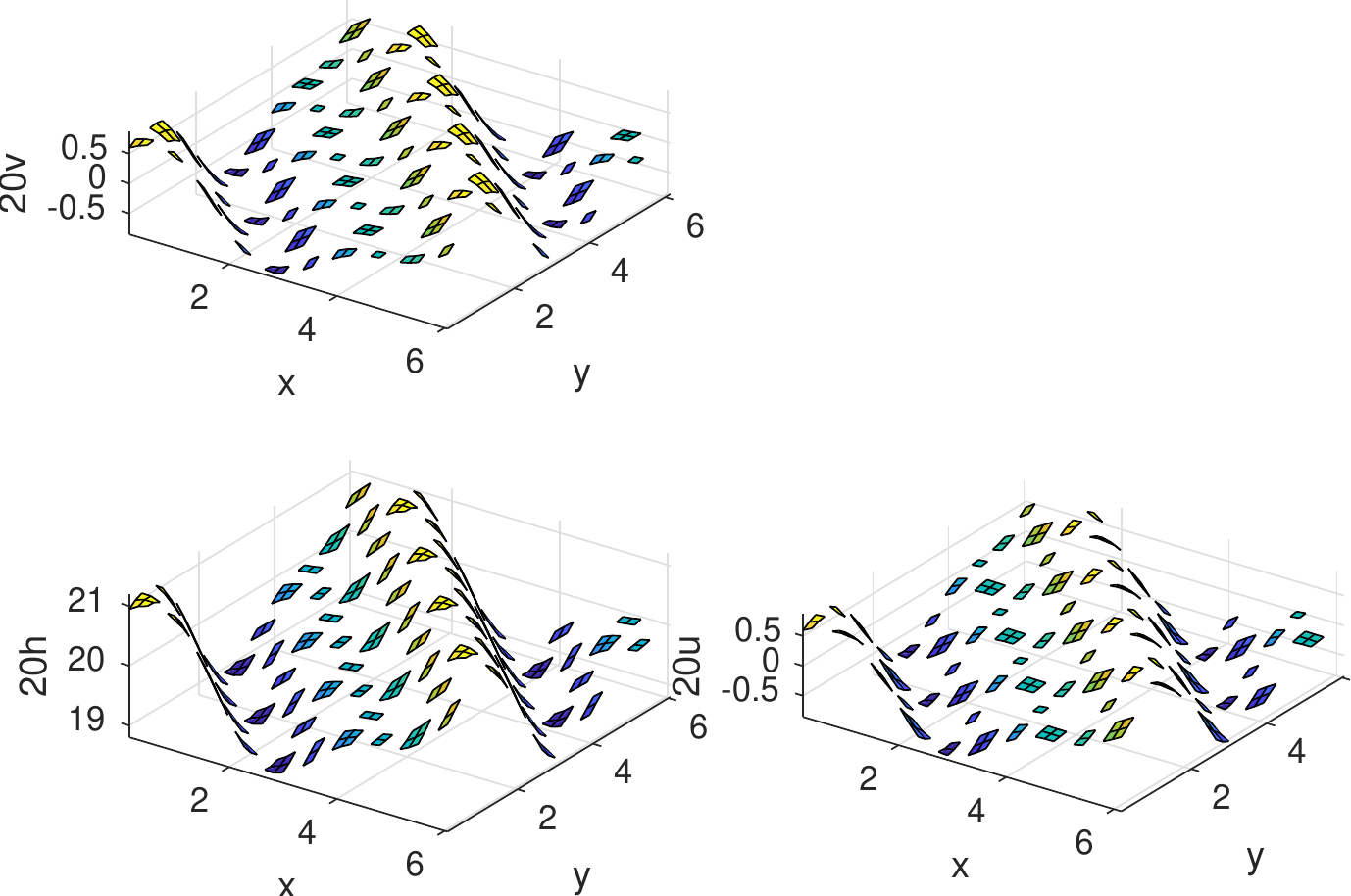}
\end{figure}%
\cref{fig:sim1t0,fig:sim1t8} 
\footnote{\url{http://www.maths.adelaide.edu.au/anthony.roberts/WWPatches/turb2DXsim1.mov} is the animation of this nonlinear wave as it progresses and distorts.}
illustrate the simulation of a progressive wave across the spatial domain.
In order to see the patches reasonably clearly we use a \(10\times10\) macro-scale grid of staggered patches (\cref{fig:patchMacroX}) of size ratio \(r=0.4\).
Each patch has a \(9\times9\) micro-scale grid with the \pde{}s~\eqref{eqs:Intro} discretised on the \(5\times5\) region in the middle of each patch.
The multiscale simulation then has \(1,475\)~evolving variables.
The initial state shown in \cref{fig:sim1t0} is of water depth \(h(x,y,0)=1+0.05\sin(x+y)\) and \(\uu(x,y,0)=\vv(x,y,0)=\frac{0.05}{\sqrt2}\sin(x+y)\).
In the simulation, the wave progresses across the domain, and gradually becomes nonlinearly\slash dispersively distorted as seen after about two periods in \cref{fig:sim1t8}.

\begin{figure}
\centering\caption{\label{fig:sim2}from the initial hump~(a), water slumps down and out~(b,c,d).  
Here use a \(22\times22\) array of \(9\times9\) patches with small size ratio \(r=0.1\), and only plot the water depth at four times.  
Since the patches are so small, this plot visually connects neighbouring patches by `ribbons' to form a `mesh'.}
\begin{tabular}{@{}ll@{}}
(a) \(t=0\) & (b) \(t=1\)\\
\includegraphics[scale=0.9]{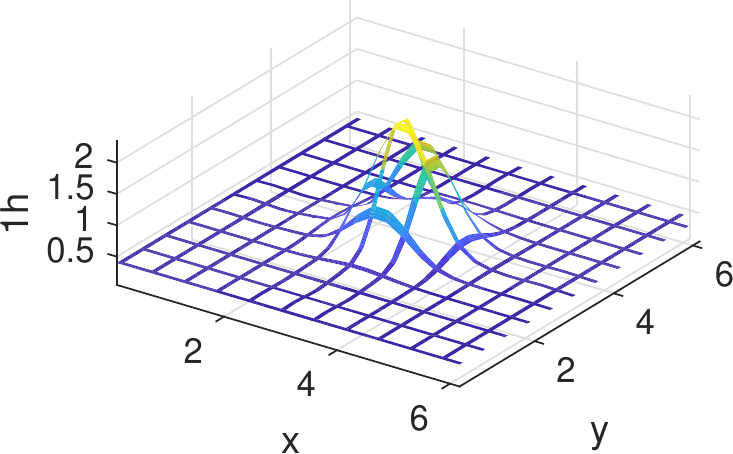}&
\includegraphics[scale=0.9]{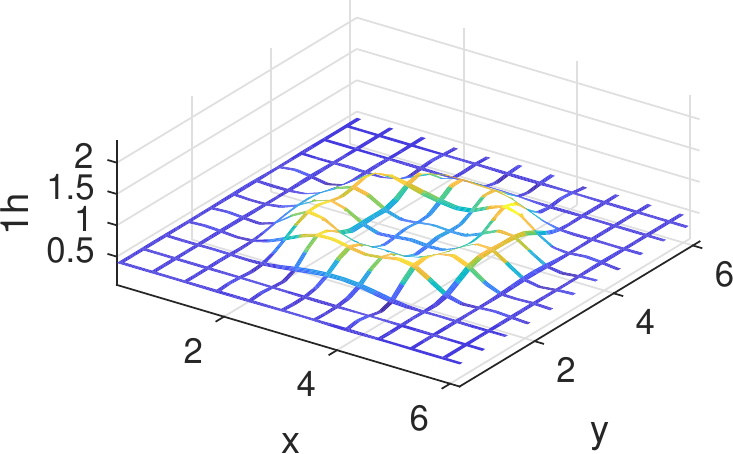}\\
(c) \(t=2\) & (d) \(t=4\)\\
\includegraphics[scale=0.9]{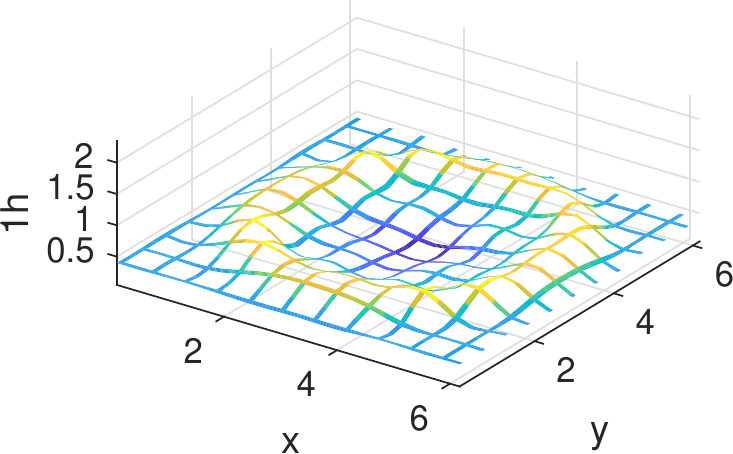}&
\includegraphics[scale=0.9]{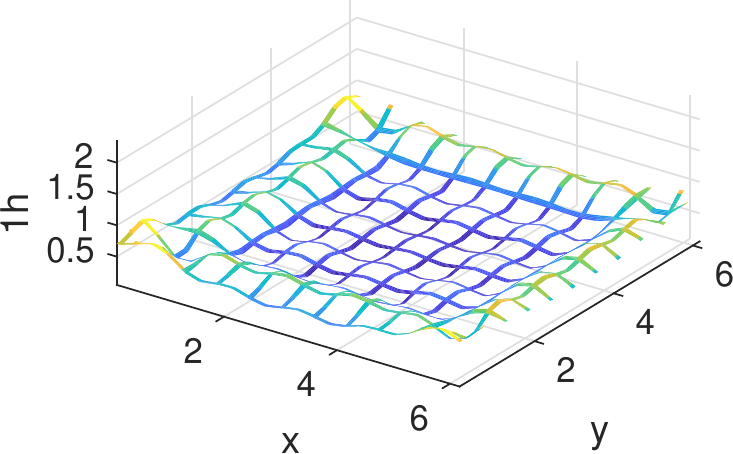}
\end{tabular}
\end{figure}%
\cref{fig:sim2}
\footnote{\url{http://www.maths.adelaide.edu.au/anthony.roberts/WWPatches/turb2DXsim2.mov} is the animation of the slumping of this water hump.} 
illustrates the simulation of the slumping of a hump of water in the spatial domain.
Here we use a \(22\times22\) macro-scale grid of staggered patches, each with  \(9\times9\) micro-grid (\cref{fig:patchMacroX}), of small size ratio \(r=0.1\).
The multiscale simulation then has \(7,139\)~evolving variables but we only compute on about~3\% of the spatial domain.
The initial state, \cref{fig:sim2}(a), is the Gaussian \(h=0.4+2\exp\big[-2(x-\pi)^2-2(y-\pi)^2\big]\) and \(\uu=\vv=0\).
The water slumps down, forming almost a radial solitary wave, \cref{fig:sim2}(b,c), until interacting with itself in the macro-periodic domain, \cref{fig:sim2}(d).

\begin{figure}
\centering\caption{\label{fig:sim3}from the initial asymmetric hump~(a), water slumps down and out~(b,c,d) in roughly two solitary waves.  
Here use a \(22\times22\) array of \(9\times9\) patches with smaller size ratio \(r=0.05\), and only plot the water depth at four times.  
Since the patches are so small, this plot visually connects neighbouring patches by `ribbons' to form a `mesh'.}
\begin{tabular}{@{}ll@{}}
(a) \(t=0\) & (b) \(t=1\)\\
\includegraphics[scale=0.9]{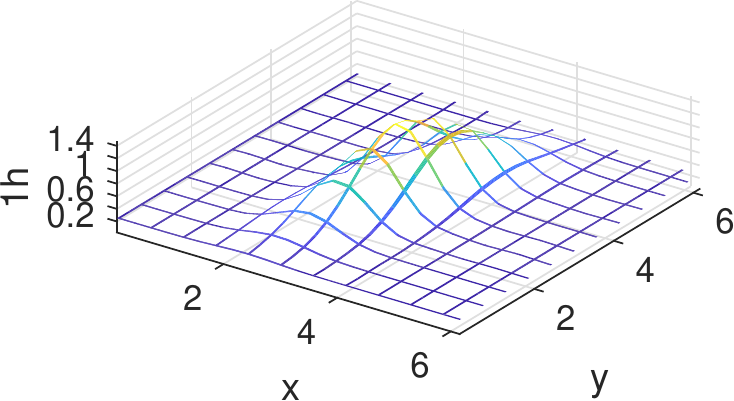}&
\includegraphics[scale=0.9]{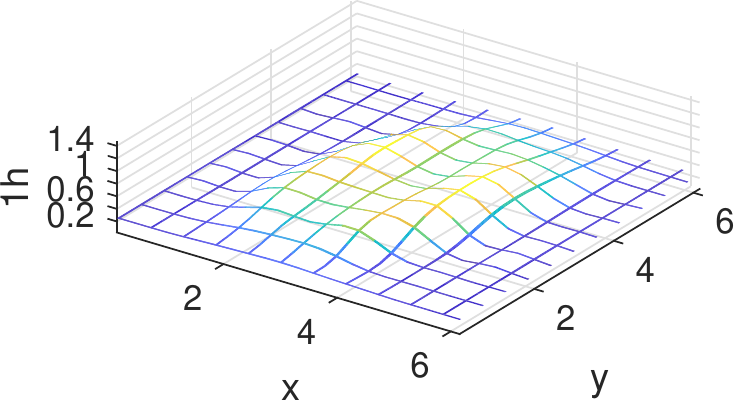}\\
(c) \(t=2\) & (d) \(t=3\)\\
\includegraphics[scale=0.9]{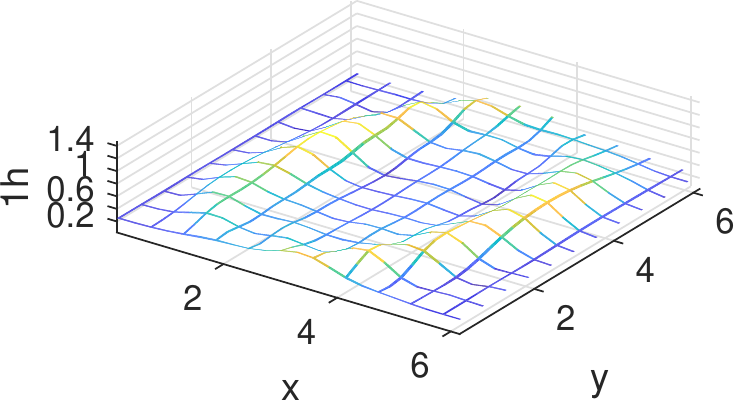}&
\includegraphics[scale=0.9]{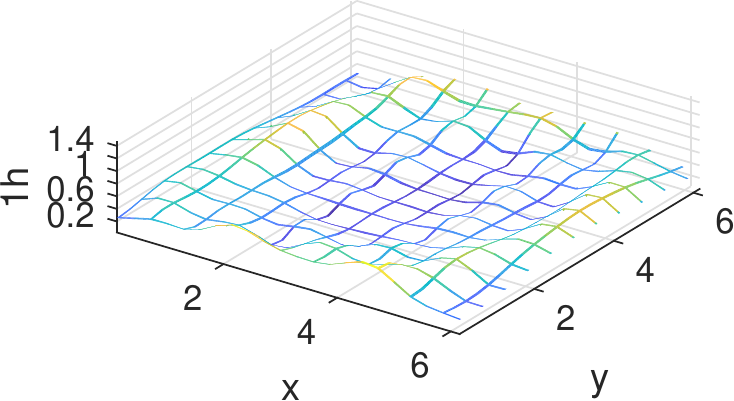}
\end{tabular}
\end{figure}

\cref{fig:sim2,fig:sim3} both plot a `mesh' of `ribbons' connecting small patches.
We do this because the patches are too small to appreciate visually.  
Both figures are for a \(22\times22\) macro-scale mesh. 
However, we only see \(11\times11\) ribbons in the plot.
The reason is that these are \emph{staggered} patches, \cref{fig:patchMacroX}, so every second potential ribbon is missing.
Further, because of the staggered patches, halfway between intersection points of the mesh lies a \(U/V\)-patch that fills in more information on the ribbons.

\cref{fig:sim3}
\footnote{\url{http://www.maths.adelaide.edu.au/anthony.roberts/WWPatches/turb2DXsim3.mov} is the animation of this asymmetric simulation.} 
similarly illustrates the simulation of the slumping of a hump of water in space but here the hump is asymmetric, the water shallower, and the patches are even smaller at size ratio \(r=0.05\).
The \(7,139\)~evolving variables of this multiscale simulation are only computed on about~0.75\% of the spatial domain.
The initial state, \cref{fig:sim3}(a), is the Gaussian \(h=0.25+1.2\exp\big[-2(x-\pi)^2+(x-\pi)(y-\pi)-\tfrac12(y-\pi)^2\big]\) and \(\uu=\vv=0\).
The water slumps down, forming two near-solitary waves, \cref{fig:sim3}(b,c,d).

\section{Conclusion}

This article lays the foundation for the accurate and efficient simulation of wave systems in a medium with complex micro-scale physics.
The `equation-free' approach \cite[e.g.,][]{Kevrekidis09a} is to compute on only small well-separated patches of space, patches that are craftily coupled to ensure that macro-scale predictions are accurate (\cref{sec:ecmclc}).
The scheme employs a macro-scale staggered grid of patches (\cref{sec:smgp}) to ensure good wave properties.
The scheme is a dynamic multiscale computational homogenisation \cite[e.g.,][]{Maier2019, Geers2010}.

\cref{sec:smgp} first addressed the canonical ideal wave \pde~\eqref{eqs:wavepdes} in 2D to establish the foundation that the scheme may be successfully applied to a wide range of wave systems in multi-D space.
\cref{sec:ntfs} then proceeded to demonstrate that it can apply to highly nonlinear wave systems by exploring properties when applied to the nonlinear Smagorinski-based model~\eqref{eqs:Intro} of turbulent shallow water flow.

The accuracy of multiscale patch schemes follows from coupling the patches via classic polynomial or spectral interpolation.
Earlier research proved this for a patch scheme a wide class of dissipative systems \cite[]{Roberts2011a}, and for 1D wave systems \cite[]{Cao2014a}.
\cref{sec:ecmclc} extended the analysis to establish consistency of the staggered patch scheme in 2D with the canonical wave \pde, albeit as yet only for the two cases of spectral and simple linear coupling.
Further research is addressing other coupling patch schemes and configurations.

The spectra of the staggered patch schemes, \cref{fig:idealevals,fig:dampevals,fig:turbevals}, show that the scheme  results in a very stiff system of differential equations to integrate in time.
When the wave problems have sufficient dissipation to damp the sub-patch waves\slash structures then stiff time integrators are effective.
However, when the micro-scale sub-patch waves are not significantly damped then stiff integrators are not effective so other integrators should be used.
Perhaps consider the so-called Projective Integration \cite[e.g.,][]{Gear02b}.

Future research is planned to develop the patch scheme to depth resolving models of shallow water flow, and then to patches of direct numerical simulation of turbulence within the water column of a patch.
Such developments require major research into the lifting and restriction of information between the macro- and micro-scales, and the automatic determination of relevant coarse variables to communicate between patches, in addition to managing the micro-scale computation within the patches.

\paragraph{Acknowledgements}
Parts of this research were funded by the Australian Research Council under grant DP150102385.  
The work of I.G.K. was partially supported by a \textsc{muri} grant by the US Army Research Office (Drs.~S. Stanton and M.~Munson).

\AtEndDocument{
\appendix
\section{Ancillary material: Matlab code for staggered patch simulation of turbulent shallow water}
\label{sec:mcptsw}

The code listed in this appendix computes the eigenvalues of \cref{sec:mtfsc}, and executes the specific simulations reported in \cref{sec:ntfs}.  
The computations of other sections were done with simpler versions of this listed code.

\subsection{Script to compute eigenvalues of turbulent shallow water model}
\begin{lstlisting}
%{ 
Use turb2DXdudt to simulate wave system on staggered patches of
staggered micro-grid.  AJR, 3 Mar 2019
%}
clear all
global p N r dX micro ordCoup

uMean=1e-1
vMean=1e-4
hMean=1e-0

% p = patch micro-grid paramater, size is (4p+1)^2 p=1,2,...
p=2
% N = even number of patches in each dirn of periodic domain
% for spectral interpolation need N/2 to be odd
N=10
% r = fraction distance a patch inner-edge is to centre next patch
r=0.1
% ordCoup = 0 for spectral and 1 for low-order interpatch coupling
ordCoup=0
% dX = space step of the inter-patch distance of patch centres
dX=2*pi/N
X=(0.5:N)*dX; Y=X; % mid-point patches
% basename for graphics files
basename='turb2DXeig';
if ordCoup, basename=[basename 'Lo']; end
% micro = index into locations of the PDEs on the micro-grid

% dx = space step of the micro-grid
dx=r*dX/(2*p-1)
% locations on axes of micro-grid points in all patches 
[x,y]=ndgrid(dx*(-2*p:2*p),X);
x=x+y; y=x;
% Maybe use double letters for 2D grids
% mesh of all locations, used for ICs
[xx,XX,yy,YY]=ndgrid(dx*(-2*p:2*p),X,dx*(-2*p:2*p),X);
xx=xx+XX; yy=yy+YY;

% micro-grid indexes inside a pxp patch where
n=4*p+1 % patch microgrid is nxn
iMid=2*p+1; % mid-point index
iie=3:2:n-2; % interior even points (relative to centre-patch)
iio=4:2:n-3; % interior odd points (empty if p=1)
io=2:2:n-1; % odd points including edges.
ie=1:2:n; % even points including edges.

% index of patch types: IH,IH are H-patches; IQ,IH are
% U-patches; IH,IQ are V-patches; and IQ,IQ are empty patches
IH=1:2:N; IQ=2:2:N;

% u(i,I,j,J) = (i,j)the microgrid value in (I,J)th patch
u=nan(n,N,n,N);
% Must set some values to find pattern of data in the 4D array
% Here set to equilibrium.
h0=@(x,y) hMean; 
u0=@(x,y) uMean; 
v0=@(x,y) vMean; 
% say h (e,H,e,H) (e,H,o,Q) (o,Q,e,H)
u(iie,IH,iie,IH)=h0(xx(iie,IH,iie,IH),yy(iie,IH,iie,IH)); 
u(iie,IH,iio,IQ)=h0(xx(iie,IH,iio,IQ),yy(iie,IH,iio,IQ));
u(iio,IQ,iie,IH)=h0(xx(iio,IQ,iie,IH),yy(iio,IQ,iie,IH));
% u field (o,H,e,H) (o,H,o,Q) (e,Q,e,H)
u(iio,IH,iie,IH)=u0(xx(iio,IH,iie,IH),yy(iio,IH,iie,IH)); 
u(iio,IH,iio,IQ)=u0(xx(iio,IH,iio,IQ),yy(iio,IH,iio,IQ));
u(iie,IQ,iie,IH)=u0(xx(iie,IQ,iie,IH),yy(iie,IQ,iie,IH));
% v field (e,H,o,H) (e,H,e,Q) (o,Q,o,H)
u(iie,IH,iio,IH)=v0(xx(iie,IH,iio,IH),yy(iie,IH,iio,IH)); 
u(iie,IH,iie,IQ)=v0(xx(iie,IH,iie,IQ),yy(iie,IH,iie,IQ));
u(iio,IQ,iio,IH)=v0(xx(iio,IQ,iio,IH),yy(iio,IQ,iio,IH));
% finally get the gather/scatter indices of micro-grid PDE points 
micro=find(~isnan(u));
nMicro=length(micro)
huv=u(micro);

%% Compute eigenvalues then plot in complex plane
small=1e-8 % perturb the surface by this much
Jac=nan(nMicro);
for j=1:nMicro
    Jac(:,j)=turb2DXdudt(1,huv+small*((1:nMicro)==j)')/small;
end
[evec,eval]=eig(Jac); eval=diag(eval);
nZeroEvals=sum(abs(eval)<1e-7)
nNearReals=sum(abs(imag(eval))<0.1)
gap = 3e-4 % adjust on case by case basis
nSlowMacro=sum( real(eval)>-gap )
nSlowSlowDecay=sum( abs(imag(eval))<0.1 & real(eval)>-gap )
nFastMicro=sum( real(eval)<-gap )
nFastWave=sum( real(eval)<-gap & imag(eval)>0.5 )
nUnstable=sum(real(eval)>1e-7)

%% plot all eigenvalues in arcsinh() transformed complex plane
xoom=@(x) asinh(x*1e6); % zoom real-part as many should be zero
figure(1),clf()
plot(xoom(real(eval)),asinh(imag(eval)),'.')
xlabel('Re\lambda'),ylabel('Im\lambda'),grid
tickx=10.^(-6:2:2); 
ticky=[0.5;1;2]*10.^(0:3);
tickx=sort([0 tickx -tickx]');
ticky=sort([0;ticky(:);-ticky(:)]);
xlim(xoom([-100 1e-4]))
set(gca,'Xtick',xoom(tickx) ...
    ,'XtickLabel',cellstr(num2str(tickx,0)) ...
    ,'XTickLabelRotation',45)
set(gca,'Ytick',asinh(ticky) ...
    ,'YtickLabel',cellstr(num2str(ticky,2)))
set(gcf,'PaperPosition',[0 0 8 10])
print('-depsc2',[basename num2str(p) 'N' num2str(N) ...
  'xr' num2str(round(10*r))  'lu' num2str(round(-log10(uMean)))  ...
  'lv' num2str(round(-log10(vMean))) ])

\end{lstlisting}

\subsection{Script to simulate turbulent shallow water model}
\begin{lstlisting}
%{ 
Use turb2DX to simulate turbulent shallow water wave system
on staggered patches of staggered micro-grid.  
AJR, 6 Mar 2019
%}
clear all
global p N r dX micro ordCoup

% p = patch micro-grid paramater, size is (4p+1)^2 p=1,2,...
p=2;
% N = even number of patches in each dirn of periodic domain
% for spectral interpolation need N/2 to be odd
N=10;
% r = fraction of distance a patch inner-edge is to centre of next patch
r=0.1;
% default scaling of the vertical axes
scal=1;
% ensure at least this much elapses between plots
dtMin=0.04; 
% default time interval for saving as eps
dtEps=1; 
% default for plotting surface alone, else all fields
plotJusth=0;
% ordCoup = 0 for spectral and 1 for low-order interpatch coupling
ordCoup=0

% Set initial conditions, and override default parameters, for
thecase=2;
switch thecase
case 1, 
    r=0.4
    scal=20; %scaling of the vertical axis
    dtEps=2; 
    tFin=8.2 
    a0=0.05
    h0=@(x,y) 1+a0*sin(x+y);
    u0=@(x,y) a0/sqrt(2)*sin(x+y);
    v0=@(x,y) a0/sqrt(2)*sin(x+y);
case 2
    N=22
    r=0.4 % big for visibility
    plotJusth=1;
    tFin=9.3
    h0=@(x,y) 0.4+2*exp(-2*(x-pi).^2-2*(y-pi).^2);
    u0=@(x,y) 0;
    v0=@(x,y) 0;
case 3
    N=22
    plotJusth=1;
    r=0.05
    tFin=3.5
    h0=@(x,y) 0.25+1.2*exp(-2*(x-pi).^2+(x-pi).*(y-pi)-0.5*(y-pi).^2);
    u0=@(x,y) 0;
    v0=@(x,y) 0;
end

% dX = space step of the inter-patch distance of patch centres
dX=2*pi/N
X=(0.5:N)*dX; Y=X; % mid-point patches
% basename for graphics files
basename='turb2DXsim';
if ordCoup, basename=[basename 'Lo']; end

% dx = space step of the micro-grid
dx=r*dX/(2*p-1)
% locations on axes of micro-grid points in all patches 
[x,y]=ndgrid(dx*(-2*p:2*p),X);
x=x+y; y=x;
% Use double letters for 2D grids.
% This mesh of all locations is used for ICs.
[xx,XX,yy,YY]=ndgrid(dx*(-2*p:2*p),X,dx*(-2*p:2*p),X);
xx=xx+XX; yy=yy+YY;

% micro-grid indexes inside a pxp patch where
n=4*p+1 % patch microgrid is nxn
iMid=2*p+1; % mid-point index
iie=3:2:n-2; % interior even points (relative to centre-patch)
iio=4:2:n-3; % interior odd points (empty if p=1)
io=2:2:n-1; % odd points including inner-edges.
ie=1:2:n; % even points including outer-edges.

% index of patch types: IH,IH are H-patches; IQ,IH are
% U-patches; IH,IQ are V-patches; and IQ,IQ are empty patches
IH=1:2:N;  IQ=2:2:N;

% u(i,I,j,J) = (i,j)the microgrid value in (I,J)th patch
u=nan(n,N,n,N);
% say h (e,H,e,H) (e,H,o,Q) (o,Q,e,H)
u(iie,IH,iie,IH)=h0(xx(iie,IH,iie,IH),yy(iie,IH,iie,IH)); 
u(iie,IH,iio,IQ)=h0(xx(iie,IH,iio,IQ),yy(iie,IH,iio,IQ));
u(iio,IQ,iie,IH)=h0(xx(iio,IQ,iie,IH),yy(iio,IQ,iie,IH));
% u field (o,H,e,H) (o,H,o,Q) (e,Q,e,H)
u(iio,IH,iie,IH)=u0(xx(iio,IH,iie,IH),yy(iio,IH,iie,IH)); 
u(iio,IH,iio,IQ)=u0(xx(iio,IH,iio,IQ),yy(iio,IH,iio,IQ));
u(iie,IQ,iie,IH)=u0(xx(iie,IQ,iie,IH),yy(iie,IQ,iie,IH));
% v field (e,H,o,H) (e,H,e,Q) (o,Q,o,H)
u(iie,IH,iio,IH)=v0(xx(iie,IH,iio,IH),yy(iie,IH,iio,IH)); 
u(iie,IH,iie,IQ)=v0(xx(iie,IH,iie,IQ),yy(iie,IH,iie,IQ));
u(iio,IQ,iio,IH)=v0(xx(iio,IQ,iio,IH),yy(iio,IQ,iio,IH));
% finally get the gather/scatter indices of micro-grid PDE points 
micro=find(~isnan(u));
nMicro=length(micro)
huv=u(micro); 

%% simulate in time, 23 takes 4x bigger steps than 45, and 3x 15s
[ts,huvs]=ode23(@turb2DXdudt,[0 tFin],huv);
nTimes=length(ts)
medianTimeStep=median(diff(ts))

%% Plot the simulation
% First, index the h and q grid points in the 2D grid
% The Nans may provide natural separators 
io=2:2:n-1; ie=1:2:n;
[ih,tmp]=ndgrid([ie n+io],0:2:N-2);
ih=reshape(ih+n*tmp,1,[]);
[iq,tmp]=ndgrid([io n+ie],0:2:N-2);
iq=reshape(iq+n*tmp,1,[]);
% Find the range of the data over all time
hmax=-Inf; hmin=+Inf;
umax=-Inf; umin=+Inf;
vmax=-Inf; vmin=+Inf;
uc=nan(n,N,n,N);%preallocate
for l=1:length(ts)
  uc(micro)=huvs(l,:);
  uu=reshape(uc,n*N,n*N); 
  hmax=max(hmax,max(max(uu(ih,ih))));
  hmin=min(hmin,min(min(uu(ih,ih))));
  umax=max(umax,max(max(uu(iq,ih))));
  umin=min(umin,min(min(uu(iq,ih))));
  vmax=max(vmax,max(max(uu(ih,iq))));
  vmin=min(vmin,min(min(uu(ih,iq))));
end
pad=0.03; % and pad range by a little
hmax=hmax+pad*(hmax-hmin); hmin=hmin-pad*(hmax-hmin); 
umax=umax+pad*(umax-umin); umin=umin-pad*(umax-umin); 
vmax=vmax+pad*(vmax-vmin); vmin=vmin-pad*(vmax-vmin);

% Prepare the new avi movie file.
vidObj = VideoWriter([basename num2str(thecase)]);
open(vidObj);

uc=nan(n,N,n,N);%preallocate
tNext=0; tEps=0; 
for l=1:length(ts)
if ts(l)>=tNext, tNext=ts(l)+dtMin;
  % choose between interpolated values and 'lines', or just patch interiors
  if r<0.3, if ordCoup, uc=stag2DXcouple(huvs(l,:));
            else uc=stag2DXspectral(huvs(l,:));
            end
  else  uc(micro)=huvs(l,:);
  end
  uu=reshape(uc,n*N,n*N); 
  if plotJusth % plot either surface alone, else all fields
    figure(1),clf()
    hs=surf(x(ih),y(ih),scal*uu(ih,ih)');
    xlabel('x'),ylabel('y'),zlabel([num2str(scal) 'h']),view(35,25)
    axis image,zlim(scal*[hmin hmax])
    if r<0.3, set(hs,'linestyle','none'); end
    set(gcf,'PaperPosition',[0 0 8 6])
  else % plot all patch fields
    figure(1),clf()
    subplot('position',[.09 .06 .4 .4])%subplot(2,2,3)
    hs=surf(x(ih),y(ih),scal*uu(ih,ih)');
    xlabel('x'),ylabel('y'),zlabel([num2str(scal) 'h']),view(35,25)
    axis image,zlim(scal*[hmin hmax])
    if r<0.3, set(hs,'linestyle','none'); end
    subplot('position',[.59 .06 .4 .4])%subplot(2,2,4)
    hs=surf(x(iq),y(ih),scal*uu(iq,ih)')
    xlabel('x'),ylabel('y'),zlabel([num2str(scal) 'u']),view(35,25)
    axis image,zlim(scal*[umin umax])
    if r<0.3, set(hs,'linestyle','none'); end
    subplot('position',[.09 .56 .4 .4])%subplot(2,2,1)
    hs=surf(x(ih),y(iq),scal*uu(ih,iq)');
    xlabel('x'),ylabel('y'),zlabel([num2str(scal) 'v']),view(35,25)
    axis image,zlim(scal*[vmin vmax])
    if r<0.3, set(hs,'linestyle','none'); end
    set(gcf,'PaperPosition',[0 0 14 10])
  end
  if ts(l)>=tEps, tEps=tEps+dtEps;
    print('-depsc',[basename num2str(thecase) ...
    't' num2str(ts(l),2) '.eps'])
  end
  title(sprintf('time =%5.2f',ts(l)))
  pause(0.01)

  % Write each frame to the file.
  writeVideo(vidObj,getframe(gcf));
end%if plot at this time
end%for-loop over time
close(vidObj); % close movie file

\end{lstlisting}

\subsection{Spectral interpolation couples patches}
\begin{lstlisting}
function [u,IH,IQ,iio,iie,dx]=stag2DXspectral(huv)
% Compute patch-edge values on staggered patches of
% staggered micro-grid when patches have edges that are two
% grid points thick.  AJR, 1 Mar 2019
% I/O: huv = vector of h,u,v values on staggered
%     patches of staggered micro-grid
% u = 4D array of huv values, with patch-edge values interpolated
% IH = H-patch indices
% IQ = U,V-patch indices, in conjunction with IH
% iio = indices of internal lattice, odd-points
% iie = indices of internal lattice, even-points
% dx = microscale lattice spacing
% Parameters
% p = patch micro-grid paramater, size is (4p+1)^2 p=1,2,...
% N = even number of patches in each dirn of periodic domain
% r = fraction of distance a patch inner-edge is to centre of next patch
% dX = space step of the inter-patch distance of patch centres
% micro = index into locations of the PDEs on the micro-grid
global p N r dX micro 

% dx = space step of the micro-grid
dx=r*dX/(2*p-1);
ro=2*p*dx/dX;  % = fraction of distance to outer-edge

% micro-grid indexes inside a nxn patch where
n=4*p+1; % patch microgrid is nxn
iMid=2*p+1; % mid-point index
iie=3:2:n-2; % interior even points (relative to the mid-patch)
iio=4:2:n-3; % interior odd points (empty when p=1)

% index of patch types: IH,IH are H-patches; IQ,IH are
% U-patches; IH,IQ are V-patches; and IQ,IQ are empty patches
IH=1:2:N; IQ=2:2:N;

% u(i,I,j,J) = (i,j)the microgrid value in (I,J)th patch
u=nan(n,N,n,N);
u(micro)=huv;

%% Interpolate spectrally to set necessary patch edge values
% Omits the corner points on all patches
% N/2, 2D Fourier coeffs of the three types of patches
cHk=fft2(squeeze(u(iMid,IH,iMid,IH))); 
cUk=fft2(squeeze(u(iMid,IQ,iMid,IH))); 
cVk=fft2(squeeze(u(iMid,IH,iMid,IQ))); 
% do we need some dXs in here? no, they cancel
kMax=(N/2-1)/2; %max wavenumber for interpolation, N/2 must be odd
% shift one patch, one macro-grid, to the right/up
k1=pi/(N/2)*(rem(kMax+(0:N/2-1),N/2)-kMax);
kr = r*k1;    kro = ro*k1;    % shift patch-half-width to the right/up
krp=(1+r)*k1; krop=(1+ro)*k1; % shift one + patch-half-width to the right/up
krm=(1-r)*k1; krom=(1-ro)*k1; % shift one - patch-half-width to the right/up

% First compute all the inner-edge gridpoints
% a dash means shift in x-dirn, undashed is in y-dirn
% a plus = shift right/up, a minus = shift left/down --- in a cell
for j=iie
  ks=(j-iMid)*dx/dX*k1;
  % interpolate U and V values to H-patches
  u(n-1,IH,j,IH)=ifft2(cUk.*exp(1i*( -krm'+ks )));
  u(2  ,IH,j,IH)=ifft2(cUk.*exp(1i*( -krp'+ks )));
  u(j,IH,n-1,IH)=ifft2(cVk.*exp(1i*( -krm +ks')));
  u(j,IH,2  ,IH)=ifft2(cVk.*exp(1i*( -krp +ks')));
  % interpolate H to U-patches
  u(n-1,IQ,j,IH)=ifft2(cHk.*exp(1i*( +krp'+ks )));
  u(2  ,IQ,j,IH)=ifft2(cHk.*exp(1i*( +krm'+ks )));
  % interpolate H to V-patches
  u(j,IH,n-1,IQ)=ifft2(cHk.*exp(1i*( +krp +ks')));
  u(j,IH,2  ,IQ)=ifft2(cHk.*exp(1i*( +krm +ks')));
end
for j=2:2:n-1
  ks=(j-iMid)*dx/dX*k1;
  % interpolate V to U-patches
  u(j,IQ,n-1,IH)=ifft2(cVk.*exp(1i*( k1'+ks'-krm )));
  u(j,IQ,2  ,IH)=ifft2(cVk.*exp(1i*( k1'+ks'-krp )));
  % interpolate U to V-patches
  u(n-1,IH,j,IQ)=ifft2(cUk.*exp(1i*( k1 +ks -krm')));
  u(2  ,IH,j,IQ)=ifft2(cUk.*exp(1i*( k1 +ks -krp')));
  % also do top/bot u of V-patches
  u(j,IH,n-1,IQ)=ifft2(cUk.*exp(1i*( krp +ks' -k1')));
  u(j,IH,2  ,IQ)=ifft2(cUk.*exp(1i*( krm +ks' -k1')));
  % and left/right v of U-patches
  u(n-1,IQ,j,IH)=ifft2(cVk.*exp(1i*( krp' +ks -k1 )));
  u(2  ,IQ,j,IH)=ifft2(cVk.*exp(1i*( krm' +ks -k1 )));
end

% Second compute all the outer-edge gridpoints
% a dash means shift in x-dirn, undashed is in y-dirn
% a plus = shift right/up, a minus = shift left/down --- in a cell
for j=2:2:n-1
  ks=(j-iMid)*dx/dX*k1;
  % interpolate U and V values to H-patches
  u(n,IH,j,IH)=ifft2(cVk.*exp(1i*( +kro'-k1+ks )));
  u(1,IH,j,IH)=ifft2(cVk.*exp(1i*( -kro'-k1+ks )));
  u(j,IH,n,IH)=ifft2(cUk.*exp(1i*( +kro-k1'+ks')));
  u(j,IH,1,IH)=ifft2(cUk.*exp(1i*( -kro-k1'+ks')));
  % interpolate H to U-patches
  u(j,IQ,n,IH)=ifft2(cHk.*exp(1i*( +k1'+ks'+kro )));
  u(j,IQ,1,IH)=ifft2(cHk.*exp(1i*( +k1'+ks'-kro )));
  % third interpolate H to V-patches
  u(n,IH,j,IQ)=ifft2(cHk.*exp(1i*( +k1+ks+kro' )));
  u(1,IH,j,IQ)=ifft2(cHk.*exp(1i*( +k1+ks-kro' )));
end
for j=1:2:n-2
  ks=(j-iMid)*dx/dX*k1;
  % interpolate top/bot H to H-patches
  u(    j,IH,n,IH)=ifft2(cHk.*exp(1i*( +ks'+kro )));
  u(n+1-j,IH,1,IH)=ifft2(cHk.*exp(1i*( -ks'-kro )));
  % and left/right H of H-patches
  u(n,IH,n+1-j,IH)=ifft2(cHk.*exp(1i*( +kro'-ks )));
  u(1,IH,    j,IH)=ifft2(cHk.*exp(1i*( -kro'+ks )));
  % interpolate top/bot U to U-patches
  u(    j,IQ,n,IH)=ifft2(cUk.*exp(1i*( +ks'+kro )));
  u(n+1-j,IQ,1,IH)=ifft2(cUk.*exp(1i*( -ks'-kro )));
  % and left/right U of U-patches
  u(n,IQ,n+1-j,IH)=ifft2(cUk.*exp(1i*( +kro'-ks )));
  u(1,IQ,    j,IH)=ifft2(cUk.*exp(1i*( -kro'+ks )));
  % interpolate top/bot V to V-patches
  u(    j,IH,n,IQ)=ifft2(cVk.*exp(1i*( +ks'+kro )));
  u(n+1-j,IH,1,IQ)=ifft2(cVk.*exp(1i*( -ks'-kro )));
  % and left/right V of V-patches
  u(n,IH,n+1-j,IQ)=ifft2(cVk.*exp(1i*( +kro'-ks )));
  u(1,IH,    j,IQ)=ifft2(cVk.*exp(1i*( -kro'+ks )));
end

\end{lstlisting}

\subsection{Micro-scale PDE discretisation within patches}
\begin{lstlisting}
function dhuvdt=turb2DXdudt(t,huv)
% Compute time derivatives of shallow water wave system on
% staggered patches of staggered micro-grid, grid0. 
% AJR, 3 Mar 2019
% I/O: t = time, ignored as autonomous
% huv = h,u,v values on staggered patches of staggered micro-grid
% dhuvdt = array of time derivatives of huv values
% Parameters (coupling has other global params)
% micro = index into locations of the PDEs on the micro-grid
% ordCoup = 0 for spectral and 1 for low-order
global micro cDrag cEddy ordCoup

% coefficients from Meng Cao & Roberts (2016)
cDrag=0.00293; % nonlinear bed drag on velocity
cEddy = 0.094; % one eddy viscosity coeff
cAduu = 1.045; % self-advection coefficient
cAduv = 1.017; % cross-advection coefficient
cGrav = 0.993; % effective gravity
qMin  = 1e-6;  % regularise turbulent dissipation
% better may be ENO-like by adding cAduu*q*dx to cEddy*q*h

% u(i,I,j,J) = (i,j)the microgrid value in (I,J)th patch
% Specify the function that couples patches by providing
% patch edge values through whatever mechanism
if ordCoup
  [ut,IH,IQ,iio,iie,dx]=stag2DXcouple(huv);
  else [ut,IH,IQ,iio,iie,dx]=stag2DXspectral(huv);
end;
io=[2 iio iio(end)+2];
ie=[1 iie iie(end)+2];
% scatter data fields into separate arrays for safety
h=nan(size(ut)); u=h; v=h; q=h;
h(ie,IH,ie,IH)=ut(ie,IH,ie,IH);
h(ie,IH,io,IQ)=ut(ie,IH,io,IQ);
h(io,IQ,ie,IH)=ut(io,IQ,ie,IH);
u(io,IH,ie,IH)=ut(io,IH,ie,IH); 
u(io,IH,io,IQ)=ut(io,IH,io,IQ);
u(ie,IQ,ie,IH)=ut(ie,IQ,ie,IH);
v(ie,IH,io,IH)=ut(ie,IH,io,IH); 
v(ie,IH,ie,IQ)=ut(ie,IH,ie,IQ);
v(io,IQ,io,IH)=ut(io,IQ,io,IH);

% q is the mean-speed field at all interior u,v points
% --- requires all u,v-edge values
q(iio,IH,iie,IH)=sqrt(u(iio,IH,iie,IH).^2 +(v(iio+1,IH,iie+1,IH) ...
+v(iio+1,IH,iie-1,IH)+v(iio-1,IH,iie+1,IH)+v(iio-1,IH,iie-1,IH)).^2/16); 
q(iio,IH,iio,IQ)=sqrt(u(iio,IH,iio,IQ).^2 +(v(iio+1,IH,iio+1,IQ) ...
+v(iio+1,IH,iio-1,IQ)+v(iio-1,IH,iio+1,IQ)+v(iio-1,IH,iio-1,IQ)).^2/16);
q(iie,IQ,iie,IH)=sqrt(u(iie,IQ,iie,IH).^2 +(v(iie+1,IQ,iie+1,IH) ...
+v(iie+1,IQ,iie-1,IH)+v(iie-1,IQ,iie+1,IH)+v(iie-1,IQ,iie-1,IH)).^2/16);
q(iie,IH,iio,IH)=sqrt(v(iie,IH,iio,IH).^2 +(u(iie+1,IH,iio+1,IH) ...
+u(iie+1,IH,iio-1,IH)+u(iie-1,IH,iio+1,IH)+u(iie-1,IH,iio-1,IH)).^2/16); 
q(iie,IH,iie,IQ)=sqrt(v(iie,IH,iie,IQ).^2 +(u(iie+1,IH,iie+1,IQ) ...
+u(iie+1,IH,iie-1,IQ)+u(iie-1,IH,iie+1,IQ)+u(iie-1,IH,iie-1,IQ)).^2/16);
q(iio,IQ,iio,IH)=sqrt(v(iio,IQ,iio,IH).^2 +(u(iio+1,IQ,iio+1,IH) ...
+u(iio+1,IQ,iio-1,IH)+u(iio-1,IQ,iio+1,IH)+u(iio-1,IQ,iio-1,IH)).^2/16);
q=q+qMin; % make bed drag more robust: reconsider micro-CFL

% For nonlinear terms we need to interpolate data to other
% micro-grid points.  First, get h at u and v points.
h(io,IH,iie,IH)=0.5*(h(io-1,IH,iie,IH)+h(io+1,IH,iie,IH)); % u-pts in H-pat
h(iie,IH,io,IH)=0.5*(h(iie,IH,io-1,IH)+h(iie,IH,io+1,IH)); % v-pts in H-pat
h(iie,IQ,iie,IH)=0.5*(h(iie-1,IQ,iie,IH)+h(iie+1,IQ,iie,IH));% u-pts in U-pat
h(io,IQ,io,IH)=0.5*(h( io,IQ,io-1,IH)+h( io,IQ,io+1,IH)); % v-pts in U-pat
h(iie,IH,iie,IQ)=0.5*(h(iie,IH,iie-1,IQ)+h(iie,IH,iie+1,IQ));% v-pts in V-pat
h(io,IH, io,IQ)=0.5*(h(io-1,IH, io,IQ)+h(io+1,IH, io,IQ)); % u-pts in V-pat
% Second get v at u-pts and u at v-pts, not outer-edge though
v(io,IH,iie,IH) =(v(io-1,IH,iie+1,IH)+v(io+1,IH,iie+1,IH) ...
                 +v(io-1,IH,iie-1,IH)+v(io+1,IH,iie-1,IH))/4;
v(iie,IQ,iie,IH)=(v(iie-1,IQ,iie+1,IH)+v(iie+1,IQ,iie+1,IH) ...
                 +v(iie-1,IQ,iie-1,IH)+v(iie+1,IQ,iie-1,IH))/4;
v(io,IH,io,IQ)  =(v(io-1,IH,io+1,IQ)+v(io+1,IH,io+1,IQ) ...
                 +v(io-1,IH,io-1,IQ)+v(io+1,IH,io-1,IQ))/4;
u(iie,IH,io,IH) =(u(iie-1,IH,io+1,IH)+u(iie+1,IH,io+1,IH) ...
                 +u(iie-1,IH,io-1,IH)+u(iie+1,IH,io-1,IH))/4;
u(io,IQ,io,IH)  =(u(io-1,IQ,io+1,IH)+u(io+1,IQ,io+1,IH) ...
                 +u(io-1,IQ,io-1,IH)+u(io+1,IQ,io-1,IH))/4;
u(iie,IH,iie,IQ)=(u(iie-1,IH,iie+1,IQ)+u(iie+1,IH,iie+1,IQ) ...
                 +u(iie-1,IH,iie-1,IQ)+u(iie+1,IH,iie-1,IQ))/4;

%% Compute time derivatives in three different patch types
ut=nan(size(ut));
% dx = space step of the micro-grid
dx2=2*dx; dy2=dx2; dx2sq=dx2^2; 

% h_t=-(hu)_x-(hv)_y at (e,H,e,H) (e,H,o,Q) (o,Q,e,H)
ut(iie,IH,iie,IH)=-(u(iie+1,IH,iie,IH).*h(iie+1,IH,iie,IH) ...
                   -u(iie-1,IH,iie,IH).*h(iie-1,IH,iie,IH))/dx2 ...
                  -(v(iie,IH,iie+1,IH).*h(iie,IH,iie+1,IH) ...
                   -v(iie,IH,iie-1,IH).*h(iie,IH,iie-1,IH))/dy2 ; 
ut(iie,IH,iio,IQ)=-(u(iie+1,IH,iio,IQ).*h(iie+1,IH,iio,IQ) ...
                   -u(iie-1,IH,iio,IQ).*h(iie-1,IH,iio,IQ))/dx2 ...
                  -(v(iie,IH,iio+1,IQ).*h(iie,IH,iio+1,IQ) ...
                   -v(iie,IH,iio-1,IQ).*h(iie,IH,iio-1,IQ))/dy2 ;
ut(iio,IQ,iie,IH)=-(u(iio+1,IQ,iie,IH).*h(iio+1,IQ,iie,IH) ...
                   -u(iio-1,IQ,iie,IH).*h(iio-1,IQ,iie,IH))/dx2 ...
                  -(v(iio,IQ,iie+1,IH).*h(iio,IQ,iie+1,IH) ...
                   -v(iio,IQ,iie-1,IH).*h(iio,IQ,iie-1,IH))/dy2 ;
                  
% u_t=-h_x-?uq/h+?qh.del2u-?u.ux-?v.uy  at (o,H,e,H) (o,H,o,Q) (e,Q,e,H)
% Eddy diffusion should be q/h.div(h^2grad u)+stuff
dels=(u(iio-2,IH,iie,IH)-2*u(iio,IH,iie,IH)+u(iio+2,IH,iie,IH))/dx2sq ...
    +(u(iio,IH,iie-2,IH)-2*u(iio,IH,iie,IH)+u(iio,IH,iie+2,IH))/dx2sq ;
ut(iio,IH,iie,IH)=-(h(iio+1,IH,iie,IH)-h(iio-1,IH,iie,IH))/dx2 ...
    -cDrag*u(iio,IH,iie,IH).*q(iio,IH,iie,IH)./h(iio,IH,iie,IH) ...
    +cEddy*dels.*q(iio,IH,iie,IH).*h(iio,IH,iie,IH) ...
    -cAduu*u(iio,IH,iie,IH).*(u(iio+2,IH,iie,IH)-u(iio-2,IH,iie,IH))/(2*dx2) ...
    -cAduv*v(iio,IH,iie,IH).*(u(iio,IH,iie+2,IH)-u(iio,IH,iie-2,IH))./(2*dy2); 
dels=(u(iio-2,IH,iio,IQ)-2*u(iio,IH,iio,IQ)+u(iio+2,IH,iio,IQ))/dx2sq ...
    +(u(iio,IH,iio-2,IQ)-2*u(iio,IH,iio,IQ)+u(iio,IH,iio+2,IQ))/dx2sq ;
ut(iio,IH,iio,IQ)=-(h(iio+1,IH,iio,IQ)-h(iio-1,IH,iio,IQ))/dx2 ...
    -cDrag*u(iio,IH,iio,IQ).*q(iio,IH,iio,IQ)./h(iio,IH,iio,IQ) ...
    +cEddy*dels.*q(iio,IH,iio,IQ).*h(iio,IH,iio,IQ)  ...
    -cAduu*u(iio,IH,iio,IQ).*(u(iio+2,IH,iio,IQ)-u(iio-2,IH,iio,IQ))/(2*dx2) ...
    -cAduv*v(iio,IH,iio,IQ).*(u(iio,IH,iio+2,IQ)-u(iio,IH,iio-2,IQ))/(2*dy2);
dels=(u(iie-2,IQ,iie,IH)-2*u(iie,IQ,iie,IH)+u(iie+2,IQ,iie,IH))/dx2sq ...
    +(u(iie,IQ,iie-2,IH)-2*u(iie,IQ,iie,IH)+u(iie,IQ,iie+2,IH))/dx2sq ;
ut(iie,IQ,iie,IH)=-(h(iie+1,IQ,iie,IH)-h(iie-1,IQ,iie,IH))/dx2 ...
    -cDrag*u(iie,IQ,iie,IH).*q(iie,IQ,iie,IH)./h(iie,IQ,iie,IH) ...
    +cEddy*dels.*q(iie,IQ,iie,IH).*h(iie,IQ,iie,IH) ...
    -cAduu*u(iie,IQ,iie,IH).*(u(iie+2,IQ,iie,IH)-u(iie-2,IQ,iie,IH))./(2*dx2) ...
    -cAduv*v(iie,IQ,iie,IH).*(u(iie,IQ,iie+2,IH)-u(iie,IQ,iie-2,IH))./(2*dy2) ;
    
% v_t=-h_y-?vq/h+?qh.del2v-?u.vx-?v.vy  at (e,H,o,H) (e,H,e,Q) (o,Q,o,H)
% Eddy diffusion should be q/h.div(h^2grad v)+stuff
dels=(v(iie-2,IH,iio,IH)-2*v(iie,IH,iio,IH)+v(iie+2,IH,iio,IH))/dx2sq ...
    +(v(iie,IH,iio-2,IH)-2*v(iie,IH,iio,IH)+v(iie,IH,iio+2,IH))/dx2sq ;
ut(iie,IH,iio,IH)=-(h(iie,IH,iio+1,IH)-h(iie,IH,iio-1,IH))/dy2 ...
    -cDrag*v(iie,IH,iio,IH).*q(iie,IH,iio,IH)./h(iie,IH,iio,IH)  ...
    +cEddy*dels.*q(iie,IH,iio,IH).*h(iie,IH,iio,IH) ...
    -cAduv*u(iie,IH,iio,IH).*(v(iie+2,IH,iio,IH)-v(iie-2,IH,iio,IH))./(2*dx2) ...
    -cAduu*v(iie,IH,iio,IH).*(v(iie,IH,iio+2,IH)-v(iie,IH,iio-2,IH))/(2*dy2); 
dels=(v(iie-2,IH,iie,IQ)-2*v(iie,IH,iie,IQ)+v(iie+2,IH,iie,IQ))/dx2sq ...
    +(v(iie,IH,iie-2,IQ)-2*v(iie,IH,iie,IQ)+v(iie,IH,iie+2,IQ))/dx2sq ;
ut(iie,IH,iie,IQ)=-(h(iie,IH,iie+1,IQ)-h(iie,IH,iie-1,IQ))/dy2 ...
    -cDrag*v(iie,IH,iie,IQ).*q(iie,IH,iie,IQ)./h(iie,IH,iie,IQ)  ...
    +cEddy*dels.*q(iie,IH,iie,IQ).*h(iie,IH,iie,IQ) ...
    -cAduv*u(iie,IH,iie,IQ).*(v(iie+2,IH,iie,IQ)-v(iie-2,IH,iie,IQ))./(2*dx2) ...
    -cAduu*v(iie,IH,iie,IQ).*(v(iie,IH,iie+2,IQ)-v(iie,IH,iie-2,IQ))./(2*dy2) ;
dels=(v(iio-2,IQ,iio,IH)-2*v(iio,IQ,iio,IH)+v(iio+2,IQ,iio,IH))/dx2sq ...
    +(v(iio,IQ,iio-2,IH)-2*v(iio,IQ,iio,IH)+v(iio,IQ,iio+2,IH))/dx2sq ;
ut(iio,IQ,iio,IH)=-(h(iio,IQ,iio+1,IH)-h(iio,IQ,iio-1,IH))/dy2 ...
    -cDrag*v(iio,IQ,iio,IH).*q(iio,IQ,iio,IH)./h(iio,IQ,iio,IH)  ...
    +cEddy*dels.*q(iio,IQ,iio,IH).*h(iio,IQ,iio,IH)  ...
    -cAduv*u(iio,IQ,iio,IH).*(v(iio+2,IQ,iio,IH)-v(iio-2,IQ,iio,IH))/(2*dx2) ...
    -cAduu*v(iio,IQ,iio,IH).*(v(iio,IQ,iio+2,IH)-v(iio,IQ,iio-2,IH))/(2*dy2);
    
% extract and reshape
dhuvdt=ut(micro);
\end{lstlisting}

\subsection{Jacobian of macro-scale modes on patches with \(n=6\) }
\begin{lstlisting}
Comment The Jacobian for the coded stag2D scheme with nearest
neighbour linear interpolation. Store in jac(6) for 6x6 patches.
Code stag2Dsv.m obtains the Jacobians by reverse engineering the
results of stag2D for various kx, ky and r.;
array jac(99);
jac(6):=(3/r)*mat(
(0,0,0,0,0,0,0,0,0,-1/2,0,0,0,0,0,-1/2,0,0,0,0,0,0,0,0,0,0,0,0,0,0
,0,0,0,0,0,(r+1)*(cos(2*ky)/4-1/4)-sin(2*ky)*((r*1i)/4+1i/4)+1/2,0
,0,0,0,0,0,0,0,0,0,0,0,0,0,(r+1)*(cos(2*kx)/4-1/4)-sin(2*kx)*((r*
1i)/4+1i/4)+1/2,0,0,0,0,0,0,0,0),
(0,0,0,0,0,0,0,0,0,0,-1/2,0,0,0,0,1/2,-1/2,0,0,0,0,0,0,0,0,0,0,0,0
,0,0,0,0,0,0,0,0,0,0,0,0,0,0,0,0,0,0,0,0,0,(r+1)*(cos(2*kx)/4-1/4)
-sin(2*kx)*((r*1i)/4+1i/4)+1/2,0,0,0,0,0,0,0,0),
(0,0,0,0,0,0,0,0,0,0,0,-1/2,0,0,0,0,1/2,0,0,0,0,0,0,0,0,0,0,0,0,0,
0,0,0,0,0,(r-1)*(cos(2*ky)/4-1/4)-sin(2*ky)*((r*1i)/4-1i/4)-1/2,0,
0,0,0,0,0,0,0,0,0,0,0,0,0,(r+1)*(cos(2*kx)/4-1/4)-sin(2*kx)*((r*1i
)/4+1i/4)+1/2,0,0,0,0,0,0,0,0),
(0,0,0,0,0,0,0,0,0,1/2,0,0,-1/2,0,0,0,0,-1/2,0,0,0,0,0,0,0,0,0,0,0
,0,0,0,0,0,0,(r+1)*(cos(2*ky)/4-1/4)-sin(2*ky)*((r*1i)/4+1i/4)+1/2
,0,0,0,0,0,0,0,0,0,0,0,0,0,0,0,0,0,0,0,0,0,0,0),
(0,0,0,0,0,0,0,0,0,0,1/2,0,0,-1/2,0,0,0,1/2,-1/2,0,0,0,0,0,0,0,0,0
,0,0,0,0,0,0,0,0,0,0,0,0,0,0,0,0,0,0,0,0,0,0,0,0,0,0,0,0,0,0,0),
(0,0,0,0,0,0,0,0,0,0,0,1/2,0,0,-1/2,0,0,0,1/2,0,0,0,0,0,0,0,0,0,0,
0,0,0,0,0,0,(r-1)*(cos(2*ky)/4-1/4)-sin(2*ky)*((r*1i)/4-1i/4)-1/2,
0,0,0,0,0,0,0,0,0,0,0,0,0,0,0,0,0,0,0,0,0,0,0),
(0,0,0,0,0,0,0,0,0,0,0,0,1/2,0,0,0,0,0,0,-1/2,0,0,0,0,0,0,0,0,0,0,
0,0,0,0,0,(r+1)*(cos(2*ky)/4-1/4)-sin(2*ky)*((r*1i)/4+1i/4)+1/2,0,
0,0,0,0,0,0,0,0,0,0,0,0,0,(r-1)*(cos(2*kx)/4-1/4)-sin(2*kx)*((r*1i
)/4-1i/4)-1/2,0,0,0,0,0,0,0,0),
(0,0,0,0,0,0,0,0,0,0,0,0,0,1/2,0,0,0,0,0,1/2,-1/2,0,0,0,0,0,0,0,0,
0,0,0,0,0,0,0,0,0,0,0,0,0,0,0,0,0,0,0,0,0,(r-1)*(cos(2*kx)/4-1/4)-
sin(2*kx)*((r*1i)/4-1i/4)-1/2,0,0,0,0,0,0,0,0),
(0,0,0,0,0,0,0,0,0,0,0,0,0,0,1/2,0,0,0,0,0,1/2,0,0,0,0,0,0,0,0,0,0
,0,0,0,0,(r-1)*(cos(2*ky)/4-1/4)-sin(2*ky)*((r*1i)/4-1i/4)-1/2,0,0
,0,0,0,0,0,0,0,0,0,0,0,0,(r-1)*(cos(2*kx)/4-1/4)-sin(2*kx)*((r*1i)
/4-1i/4)-1/2,0,0,0,0,0,0,0,0),
(1/2,0,0,-1/2,0,0,0,0,0,0,0,0,0,0,0,0,0,0,0,0,0,0,0,0,0,0,0,0,0,0,
0,0,0,0,0,0,0,0,0,0,0,0,0,0,0,0,0,0,0,0,0,0,0,0,0,0,0,0,0),
(0,1/2,0,0,-1/2,0,0,0,0,0,0,0,0,0,0,0,0,0,0,0,0,0,0,0,0,0,0,0,0,0,
0,0,0,0,0,0,0,0,0,0,0,0,0,0,0,0,0,0,0,0,0,0,0,0,0,0,0,0,0),
(0,0,1/2,0,0,-1/2,0,0,0,0,0,0,0,0,0,0,0,0,0,0,0,0,0,0,0,0,0,0,0,0,
0,0,0,0,0,0,0,0,0,0,0,0,0,0,0,0,0,0,0,0,0,0,0,0,0,0,0,0,0),
(0,0,0,1/2,0,0,-1/2,0,0,0,0,0,0,0,0,0,0,0,0,0,0,0,0,0,0,0,0,0,0,0,
0,0,0,0,0,0,0,0,0,0,0,0,0,0,0,0,0,0,0,0,0,0,0,0,0,0,0,0,0),
(0,0,0,0,1/2,0,0,-1/2,0,0,0,0,0,0,0,0,0,0,0,0,0,0,0,0,0,0,0,0,0,0,
0,0,0,0,0,0,0,0,0,0,0,0,0,0,0,0,0,0,0,0,0,0,0,0,0,0,0,0,0),
(0,0,0,0,0,1/2,0,0,-1/2,0,0,0,0,0,0,0,0,0,0,0,0,0,0,0,0,0,0,0,0,0,
0,0,0,0,0,0,0,0,0,0,0,0,0,0,0,0,0,0,0,0,0,0,0,0,0,0,0,0,0),
(1/2,-1/2,0,0,0,0,0,0,0,0,0,0,0,0,0,0,0,0,0,0,0,0,0,0,0,0,0,0,0,0,
0,0,0,0,0,0,0,0,0,0,0,0,0,0,0,0,0,0,0,0,0,0,0,0,0,0,0,0,0),
(0,1/2,-1/2,0,0,0,0,0,0,0,0,0,0,0,0,0,0,0,0,0,0,0,0,0,0,0,0,0,0,0,
0,0,0,0,0,0,0,0,0,0,0,0,0,0,0,0,0,0,0,0,0,0,0,0,0,0,0,0,0),
(0,0,0,1/2,-1/2,0,0,0,0,0,0,0,0,0,0,0,0,0,0,0,0,0,0,0,0,0,0,0,0,0,
0,0,0,0,0,0,0,0,0,0,0,0,0,0,0,0,0,0,0,0,0,0,0,0,0,0,0,0,0),
(0,0,0,0,1/2,-1/2,0,0,0,0,0,0,0,0,0,0,0,0,0,0,0,0,0,0,0,0,0,0,0,0,
0,0,0,0,0,0,0,0,0,0,0,0,0,0,0,0,0,0,0,0,0,0,0,0,0,0,0,0,0),
(0,0,0,0,0,0,1/2,-1/2,0,0,0,0,0,0,0,0,0,0,0,0,0,0,0,0,0,0,0,0,0,0,
0,0,0,0,0,0,0,0,0,0,0,0,0,0,0,0,0,0,0,0,0,0,0,0,0,0,0,0,0),
(0,0,0,0,0,0,0,1/2,-1/2,0,0,0,0,0,0,0,0,0,0,0,0,0,0,0,0,0,0,0,0,0,
0,0,0,0,0,0,0,0,0,0,0,0,0,0,0,0,0,0,0,0,0,0,0,0,0,0,0,0,0),
(0,0,0,0,0,0,0,0,0,0,0,0,0,0,0,0,0,0,0,0,0,0,0,0,0,0,0,-1/2,0,0,0,
1/2,-1/2,0,0,0,0,0,0,0,0,0,0,0,0,0,0,0,0,0,-((cos(2*ky)+sin(2*ky)*
1i+1)*(r-cos(2*kx)+sin(2*kx)*1i+r*sin(2*kx)*1i-r*cos(2*kx)-1))/8,0
,0,0,0,0,0,0,0),
(0,0,0,0,0,0,0,0,0,0,0,0,0,0,0,0,0,0,0,0,0,0,0,0,0,0,0,0,-1/2,0,0,
0,1/2,-1/2,0,0,0,0,0,0,0,0,0,0,0,0,0,0,0,0,-((cos(2*ky)+sin(2*ky)*
1i+1)*(r-cos(2*kx)+sin(2*kx)*1i+r*sin(2*kx)*1i-r*cos(2*kx)-1))/8,0
,0,0,0,0,0,0,0),
(0,0,0,0,0,0,0,0,0,0,0,0,0,0,0,0,0,0,0,0,0,0,0,0,0,0,0,1/2,0,-1/2,
0,0,0,0,1/2,-1/2,0,0,0,0,0,0,0,0,0,0,0,0,0,0,0,0,0,0,0,0,0,0,0),
(0,0,0,0,0,0,0,0,0,0,0,0,0,0,0,0,0,0,0,0,0,0,0,0,0,0,0,0,1/2,0,-1/
2,0,0,0,0,1/2,-1/2,0,0,0,0,0,0,0,0,0,0,0,0,0,0,0,0,0,0,0,0,0,0),
(0,0,0,0,0,0,0,0,0,0,0,0,0,0,0,0,0,0,0,0,0,0,0,0,0,0,0,0,0,1/2,0,0
,0,0,0,0,0,1/2,-1/2,0,0,0,0,0,0,0,0,0,0,0,-((cos(2*ky)+sin(2*ky)*
1i+1)*(r+cos(2*kx)-sin(2*kx)*1i+r*sin(2*kx)*1i-r*cos(2*kx)+1))/8,0
,0,0,0,0,0,0,0),
(0,0,0,0,0,0,0,0,0,0,0,0,0,0,0,0,0,0,0,0,0,0,0,0,0,0,0,0,0,0,1/2,0
,0,0,0,0,0,0,1/2,-1/2,0,0,0,0,0,0,0,0,0,0,-((cos(2*ky)+sin(2*ky)*
1i+1)*(r+cos(2*kx)-sin(2*kx)*1i+r*sin(2*kx)*1i-r*cos(2*kx)+1))/8,0
,0,0,0,0,0,0,0),
(0,0,0,0,0,0,0,0,0,0,0,0,0,0,0,0,0,0,0,0,0,1/2,0,-1/2,0,0,0,0,0,0,
0,0,0,0,0,0,0,0,0,0,0,0,0,0,0,0,0,0,0,0,0,0,0,0,0,0,0,0,0),
(0,0,0,0,0,0,0,0,0,0,0,0,0,0,0,0,0,0,0,0,0,0,1/2,0,-1/2,0,0,0,0,0,
0,0,0,0,0,0,0,0,0,0,0,0,0,0,0,0,0,0,0,0,0,0,0,0,0,0,0,0,0),
(0,0,0,0,0,0,0,0,0,0,0,0,0,0,0,0,0,0,0,0,0,0,0,1/2,0,-1/2,0,0,0,0,
0,0,0,0,0,0,0,0,0,0,0,0,0,0,0,0,0,0,0,0,0,0,0,0,0,0,0,0,0),
(0,0,0,0,0,0,0,0,0,0,0,0,0,0,0,0,0,0,0,0,0,0,0,0,1/2,0,-1/2,0,0,0,
0,0,0,0,0,0,0,0,0,0,0,0,0,0,0,0,0,0,0,0,0,0,0,0,0,0,0,0,0),
(0,0,0,0,1/2-sin(2*ky)*((r*1i)/4-1i/4)-(r-1)*(cos(2*ky)/4-1/4),0,0
,0,0,0,0,0,0,0,0,0,0,0,0,0,0,-1/2,0,0,0,0,0,0,0,0,0,0,0,0,0,0,0,0,
0,0,0,0,0,0,0,0,0,0,0,0,0,0,0,0,0,0,0,0,0),
(0,0,0,0,0,0,0,0,0,0,0,0,0,0,0,0,0,0,0,0,0,1/2,-1/2,0,0,0,0,0,0,0,
0,0,0,0,0,0,0,0,0,0,0,0,0,0,0,0,0,0,0,0,0,0,0,0,0,0,0,0,0),
(0,0,0,0,-(r+1)*(cos(2*ky)/4-1/4)-sin(2*ky)*((r*1i)/4+1i/4)-1/2,0,
0,0,0,0,0,0,0,0,0,0,0,0,0,0,0,0,1/2,0,0,0,0,0,0,0,0,0,0,0,0,0,0,0,
0,0,0,0,0,0,0,0,0,0,0,0,0,0,0,0,0,0,0,0,0),
(0,0,0,0,1/2-sin(2*ky)*((r*1i)/4-1i/4)-(r-1)*(cos(2*ky)/4-1/4),0,0
,0,0,0,0,0,0,0,0,0,0,0,0,0,0,0,0,-1/2,0,0,0,0,0,0,0,0,0,0,0,0,0,0,
0,0,0,0,0,0,0,0,0,0,0,0,0,0,0,0,0,0,0,0,0),
(0,0,0,0,0,0,0,0,0,0,0,0,0,0,0,0,0,0,0,0,0,0,0,1/2,-1/2,0,0,0,0,0,
0,0,0,0,0,0,0,0,0,0,0,0,0,0,0,0,0,0,0,0,0,0,0,0,0,0,0,0,0),
(0,0,0,0,-(r+1)*(cos(2*ky)/4-1/4)-sin(2*ky)*((r*1i)/4+1i/4)-1/2,0,
0,0,0,0,0,0,0,0,0,0,0,0,0,0,0,0,0,0,1/2,0,0,0,0,0,0,0,0,0,0,0,0,0,
0,0,0,0,0,0,0,0,0,0,0,0,0,0,0,0,0,0,0,0,0),
(0,0,0,0,1/2-sin(2*ky)*((r*1i)/4-1i/4)-(r-1)*(cos(2*ky)/4-1/4),0,0
,0,0,0,0,0,0,0,0,0,0,0,0,0,0,0,0,0,0,-1/2,0,0,0,0,0,0,0,0,0,0,0,0,
0,0,0,0,0,0,0,0,0,0,0,0,0,0,0,0,0,0,0,0,0),
(0,0,0,0,0,0,0,0,0,0,0,0,0,0,0,0,0,0,0,0,0,0,0,0,0,1/2,-1/2,0,0,0,
0,0,0,0,0,0,0,0,0,0,0,0,0,0,0,0,0,0,0,0,0,0,0,0,0,0,0,0,0),
(0,0,0,0,-(r+1)*(cos(2*ky)/4-1/4)-sin(2*ky)*((r*1i)/4+1i/4)-1/2,0,
0,0,0,0,0,0,0,0,0,0,0,0,0,0,0,0,0,0,0,0,1/2,0,0,0,0,0,0,0,0,0,0,0,
0,0,0,0,0,0,0,0,0,0,0,0,0,0,0,0,0,0,0,0,0),
(0,0,0,0,0,0,0,0,0,0,0,0,0,0,0,0,0,0,0,0,0,0,0,0,0,0,0,0,0,0,0,0,0
,0,0,-((cos(2*kx)+sin(2*kx)*1i+1)*(sin(2*ky)+cos(2*ky)*1i-r*1i+r*
sin(2*ky)+r*cos(2*ky)*1i+1i)*1i)/8,0,0,0,0,0,0,0,0,0,0,1/2,0,0,-1/
2,0,0,0,0,0,-1/2,0,0,0),
(0,0,0,0,0,0,0,0,0,0,0,0,0,0,0,0,0,0,0,0,0,0,0,0,0,0,0,0,0,0,0,0,0
,0,0,0,0,0,0,0,0,0,0,0,0,0,0,1/2,0,0,-1/2,0,0,0,0,1/2,-1/2,0,0),
(0,0,0,0,0,0,0,0,0,0,0,0,0,0,0,0,0,0,0,0,0,0,0,0,0,0,0,0,0,0,0,0,0
,0,0,((cos(2*kx)+sin(2*kx)*1i+1)*(r*1i+cos(2*ky)*1i+sin(2*ky)-r*
sin(2*ky)-r*cos(2*ky)*1i+1i)*1i)/8,0,0,0,0,0,0,0,0,0,0,0,0,1/2,0,0
,-1/2,0,0,0,0,1/2,0,0),
(0,0,0,0,0,0,0,0,0,0,0,0,0,0,0,0,0,0,0,0,0,0,0,0,0,0,0,0,0,0,0,0,0
,0,0,-((cos(2*kx)+sin(2*kx)*1i+1)*(sin(2*ky)+cos(2*ky)*1i-r*1i+r*
sin(2*ky)+r*cos(2*ky)*1i+1i)*1i)/8,0,0,0,0,0,0,0,0,0,0,0,0,0,1/2,0
,0,-1/2,0,0,0,0,-1/2,0),
(0,0,0,0,0,0,0,0,0,0,0,0,0,0,0,0,0,0,0,0,0,0,0,0,0,0,0,0,0,0,0,0,0
,0,0,0,0,0,0,0,0,0,0,0,0,0,0,0,0,0,1/2,0,0,-1/2,0,0,0,1/2,-1/2),
(0,0,0,0,0,0,0,0,0,0,0,0,0,0,0,0,0,0,0,0,0,0,0,0,0,0,0,0,0,0,0,0,0
,0,0,((cos(2*kx)+sin(2*kx)*1i+1)*(r*1i+cos(2*ky)*1i+sin(2*ky)-r*
sin(2*ky)-r*cos(2*ky)*1i+1i)*1i)/8,0,0,0,0,0,0,0,0,0,0,0,0,0,0,0,1
/2,0,0,-1/2,0,0,0,1/2),
(0,0,0,0,1/2-sin(2*kx)*((r*1i)/4-1i/4)-(r-1)*(cos(2*kx)/4-1/4),0,0
,0,0,0,0,0,0,0,0,0,0,0,0,0,0,0,0,0,0,0,0,0,0,0,0,0,0,0,0,0,0,0,0,0
,-1/2,0,0,0,0,0,0,0,0,0,0,0,0,0,0,0,0,0,0),
(0,0,0,0,1/2-sin(2*kx)*((r*1i)/4-1i/4)-(r-1)*(cos(2*kx)/4-1/4),0,0
,0,0,0,0,0,0,0,0,0,0,0,0,0,0,0,0,0,0,0,0,0,0,0,0,0,0,0,0,0,0,0,0,0
,0,-1/2,0,0,0,0,0,0,0,0,0,0,0,0,0,0,0,0,0),
(0,0,0,0,1/2-sin(2*kx)*((r*1i)/4-1i/4)-(r-1)*(cos(2*kx)/4-1/4),0,0
,0,0,0,0,0,0,0,0,0,0,0,0,0,0,0,0,0,0,0,0,0,0,0,0,0,0,0,0,0,0,0,0,0
,0,0,-1/2,0,0,0,0,0,0,0,0,0,0,0,0,0,0,0,0),
(0,0,0,0,0,0,0,0,0,0,0,0,0,0,0,0,0,0,0,0,0,0,0,0,0,0,0,0,0,0,0,0,0
,0,0,0,0,0,0,0,1/2,0,0,-1/2,0,0,0,0,0,0,0,0,0,0,0,0,0,0,0),
(0,0,0,0,0,0,0,0,0,0,0,0,0,0,0,0,0,0,0,0,0,0,0,0,0,0,0,0,0,0,0,0,0
,0,0,0,0,0,0,0,0,1/2,0,0,-1/2,0,0,0,0,0,0,0,0,0,0,0,0,0,0),
(0,0,0,0,0,0,0,0,0,0,0,0,0,0,0,0,0,0,0,0,0,0,0,0,0,0,0,0,0,0,0,0,0
,0,0,0,0,0,0,0,0,0,1/2,0,0,-1/2,0,0,0,0,0,0,0,0,0,0,0,0,0),
(0,0,0,0,-(r+1)*(cos(2*kx)/4-1/4)-sin(2*kx)*((r*1i)/4+1i/4)-1/2,0,
0,0,0,0,0,0,0,0,0,0,0,0,0,0,0,0,0,0,0,0,0,0,0,0,0,0,0,0,0,0,0,0,0,
0,0,0,0,1/2,0,0,0,0,0,0,0,0,0,0,0,0,0,0,0),
(0,0,0,0,-(r+1)*(cos(2*kx)/4-1/4)-sin(2*kx)*((r*1i)/4+1i/4)-1/2,0,
0,0,0,0,0,0,0,0,0,0,0,0,0,0,0,0,0,0,0,0,0,0,0,0,0,0,0,0,0,0,0,0,0,
0,0,0,0,0,1/2,0,0,0,0,0,0,0,0,0,0,0,0,0,0),
(0,0,0,0,-(r+1)*(cos(2*kx)/4-1/4)-sin(2*kx)*((r*1i)/4+1i/4)-1/2,0,
0,0,0,0,0,0,0,0,0,0,0,0,0,0,0,0,0,0,0,0,0,0,0,0,0,0,0,0,0,0,0,0,0,
0,0,0,0,0,0,1/2,0,0,0,0,0,0,0,0,0,0,0,0,0),
(0,0,0,0,0,0,0,0,0,0,0,0,0,0,0,0,0,0,0,0,0,0,0,0,0,0,0,0,0,0,0,0,0
,0,0,0,0,0,0,0,1/2,-1/2,0,0,0,0,0,0,0,0,0,0,0,0,0,0,0,0,0),
(0,0,0,0,0,0,0,0,0,0,0,0,0,0,0,0,0,0,0,0,0,0,0,0,0,0,0,0,0,0,0,0,0
,0,0,0,0,0,0,0,0,1/2,-1/2,0,0,0,0,0,0,0,0,0,0,0,0,0,0,0,0),
(0,0,0,0,0,0,0,0,0,0,0,0,0,0,0,0,0,0,0,0,0,0,0,0,0,0,0,0,0,0,0,0,0
,0,0,0,0,0,0,0,0,0,0,1/2,-1/2,0,0,0,0,0,0,0,0,0,0,0,0,0,0),
(0,0,0,0,0,0,0,0,0,0,0,0,0,0,0,0,0,0,0,0,0,0,0,0,0,0,0,0,0,0,0,0,0
,0,0,0,0,0,0,0,0,0,0,0,1/2,-1/2,0,0,0,0,0,0,0,0,0,0,0,0,0))$
end;
\end{lstlisting}

\subsection{Approximate slow eigenspace of macro-scale modes on patches}
\begin{lstlisting}
Comment Computer Algebra (Reduce) to construct a systematic
approximation to the macroscale eigenvalues of the Jacobians for
patches applied to the classic 2D wave PDEs.  Use Jacobians stored
in file jacs.txt.  Constructs the first few orders of a series
expansion for the eigenvalues and eigenfunctions corresponding to
the zero eigenvalues that occurs at r=kx=ky=0.  To cover both
small r and small kx,ky posit an artificial small parameter,
notionally e, and expand in that.  Solve
(J0+e*J1)(V0+e*V1+e^2*V2)=(V0+e*V1+e^2*V2)(e*L1+e^2*L2)
Expand and group powers of e to get eqns.  The trick is to
construct the expansion with three macroscale modes separated from
all the others.  The case n=10 and o=5 takes 7.9 secs to execute
on my Mac.   AJR, 2 Aug 2018;

o:=5; % compute the eig-expansion to this order, errors O(o+1)
n:=6; % each patch is (n+1)x(n+1) micro-lattice

halfn:=n/2;   % 1,3,5,...
nJ:=halfn^2+2*(n-1)^2; % 3,59,187,...
in "jacs.txt"$% load all Jacobians constructed so far
jj:=jac(n)*(r/halfn)$ % select and scale the Jacobian to analyse

load_package linalg;
matrixproc cctp(a); hermitian_tp(a);
procedure matSize(a); {row_dim(a),column_dim(a)};
matrixproc ones(m,n); extend(mat((1)),m-1,n-1,1);
matrixproc zeros(m,n); extend(mat((0)),m-1,n-1,0);

sizeJacobian:=matSize(jj);
if matSize(jj) neq {nJ,nJ} then rederr("Jacobian appears wrong size");

write "set how many micro-grid pts there are of each type";
nhh:=halfn^2;        % 1,9,25,...
nuh:=halfn*(halfn-1);% 0,6,20,...
nuv:=(halfn-1)^2;    % 0,4,16,...
n0ev:=(3*n^2-8*n+16)/4; % 3,19,59,...
zer0:=zeros(n0ev,n0ev)$   % useful matrix
id0:=make_identity(n0ev)$ % also useful

jj0:=sub({kx=0,ky=0},jj)$
jj1:=jj-jj0$
write "set three base macro-scale eigenfunctions";
vh:=if n=2 then tp mat((1,0,0)) else matrix_stack({
    ones(nhh,1),zeros(nuh,1),zeros(nuh,1),
    ones(nuh,1),zeros(nuv,1),zeros(nhh,1),
    ones(nuh,1),zeros(nhh,1),zeros(nuv,1) })$
vu:=if n=2 then tp mat((0,1,0)) else matrix_stack({
    zeros(nhh,1),ones(nuh,1),zeros(nuh,1),
    zeros(nuh,1),ones(nuv,1),zeros(nhh,1),
    zeros(nuh,1),ones(nhh,1),zeros(nuv,1) })$
vv:=ones(nJ,1)-vu-vh$
write "make basis V for null space from them and others";
vs:=nullspace(matrix_stack({jj0,tp vh,tp vu,tp vv}))$
if length(vs) neq n0ev-3 then rederr("nullspace V0 appears wrong size");
if length(vs)>0 then <<
    vv0:=matrix_augment(for i:=1:length(vs) collect part(vs,i))$
    vv0:=matrix_augment({vh,vu,vv,vv0})$ >>
else vv0:=matrix_augment({vh,vu,vv});
if jj0*vv0 neq zeros(nJ,n0ev) then rederr("V0 not a nullspace");;
write "for adjoint nullspace, get basis orthogonal to V0";
zs:=nullspace(cctp jj0)$
if length(zs) neq n0ev then rederr("nullspace Z appears wrong size");
zz:=matrix_augment(for i:=1:n0ev collect part(zs,i))$
zz:=zz/(cctp vv0*zz)$
if cctp zz*vv0 neq id0 then rederr("Z and V0 not orthogonal");
write "Macro amplitude defn requires change to these grid-values";
% AmpO=full(sparse([5 51 36],1:3,1,59,3))
AmpO:=zeros(nJ,3)$
AmpO((nhh+1)/2,1):=
AmpO(nJ-nuv-(nhh-1)/2,2):=
AmpO((3*nhh+1)/2+3*nuh+nuv,3):=1$

write "Solve first order J0*V1+J1*V0=V0*L1";
write "compute initial VL";
rhs0:=matrix_stack({jj1*vv0,zer0})$
jvz0:=block_matrix(2,2,{-jj0,vv0,cctp zz,zer0})$
vl:=(1/jvz0)*rhs0$
ivv:=(for i:=1:nJ collect i)$ %indices
ill:=(for i:=nJ+1:nJ+n0ev collect i)$ %indices
ll1:=stack_rows(vl,ill)$
write "Need to zero most of the first row---tricky stuff";
ks:={kx,ky};
pps:=id0$
for i:=4:n0ev do if ll1(1,i) neq 0 then for k:=1:2 do begin
    write tanq:=df(ll1(1,i),part(ks,k))/df(ll1(1,k+1),part(ks,k));
    pp:=id0;  pp(k+1,i):=-tanq;
    ll1:=(1/pp)*ll1*pp;
    pps:=pps*pp;
end;

write "Hereafter use new nullspace bases V0*Ps and Z/Ps^T";
clear vv;
array vv(o),ll(o),llmac(o),llfac(o),eig(o);
vv(0):=vv0*pps$
zz:=zz/(cctp pps)$
if (cctp zz)*vv(0) neq id0 then rederr("not identity");
factor r;
nocos:={cos(~a)^2=>1-sin(a)^2};
trigs:={cos(~a)^2=>1-sin(a)^2
   , sin(2*~a)=>2*sin(a)*cos(a)
   , cos(2*~a)=>1-2*sin(a)^2 };

write "Re-solve first order J0*V1+J1*V0=V0*L1 to confirm,
then proceed to the higher orders";
il0:=(for i:=4:n0ev collect i)$ %indices
zzt:=if il0={} then AmpO
    else matrix_augment({AmpO,sub_matrix(zz,ivv,il0)})$
invjvz0:=1/block_matrix(2,2,{-jj0,vv(0),cctp zzt,zer0})$

write "Following LL0 appears to factor all top-left 3x3 submatrices.";
ll0:=mat(
  (0,-i*sin(kx)*(cos(kx)-i*sin(kx)),-i*sin(ky)*(cos(ky)-i*sin(ky)))
  ,(-i*sin(kx)*(cos(kx)+i*sin(kx)),0,0)
  ,(-i*sin(ky)*(cos(ky)+i*sin(ky)),0,0) );
for k:=1:o do begin
  rhs0:=jj1*vv(k-1)$
  for j:=1:k-1 do rhs0:=rhs0-vv(k-j)*ll(j)$
  rhs0:=matrix_stack({rhs0,zer0})$
  vl:=(invjvz0*rhs0 where nocos)$
  ll(k):=stack_rows(vl,ill);
  vv(k):=stack_rows(vl,ivv)$
  write llmac(k):=(sub_matrix(ll(k),{1,2,3},{1,2,3})
       where trigs);
  tmp:=llmac(k)*halfn/r$
  write llfac(k):=trigsimp( tmp(1,2)/sin(kx)*(i*cos(kx)-sin(kx)) );  
  tmp:=llmac(k)-r/halfn*llfac(k)*ll0;
  if tmp neq zeros(3,3) then rederr("factorisation error");
end;

showtime;
end;
\end{lstlisting}
}

\end{document}

%% file: microStaggerGrid.tex
\ifcase2
\or\def\N{3}\def\Nm{2}\def\Np{4}\def\NN{6}\def\NNp{7}
\or\def\N{5}\def\Nm{4}\def\Np{6}\def\NN{10}\def\NNp{11}
\fi%
\setlength{\unitlength}{1.8ex}
\begin{picture}(11,11)
\setcounter{j}{0}
\multiput(0.2,0.5)(0,12)1{\stepcounter{j}
\setcounter{i}{0}
\multiput(0,0)(12,0)1{\stepcounter{i}
\color{green}
\ifodd\value{i}%
\multiput(0,0)(0,1){\NNp}{\line(1,0){\NN}}%
\multiput(0,0)(1,0){\NNp}{\line(0,1){\NN}}%
\else\ifodd\value{j}%
\multiput(0,0)(0,1){\NNp}{\line(1,0){\NN}}%
\multiput(0,0)(1,0){\NNp}{\line(0,1){\NN}}%
\fi\fi
\def\cc{white}\def\cx{yellow}\def\cy{yellow}%
\ifodd\value{j}%
  \ifodd\value{i}
    \def\cc{black}\def\cx{cyan}\def\cy{red}%
    \def\ha{1}\def\hr{\N}%
    \def\hA{1}\def\hR{\N}%
  \else
    \def\cc{cyan}\def\cx{black}\def\cy{red}%
    \def\ha{1}\def\hr{\N}%
    \def\hA{0}\def\hR{\Np}
  \fi%
\else%
  \ifodd\value{i}
    \def\cc{red}\def\cx{cyan}\def\cy{black}%
    \def\ha{0}\def\hr{\Np}
    \def\hA{1}\def\hR{\N}%
  \else
    \def\ha{0}\def\hr{0}%
    \def\hA{0}\def\hR{0}%
  \fi%
\fi%
\color{\cc}
\multiput(1,1)(2,0)\N{\multiput(0,0)(0,2)\N{\circle*{0.5}}}
\color{\cx}
\multiput(0,\ha)(2,0)\Np{\multiput(0,0)(0,2)\hr{\circle*{0.5}}}
\color{\cy}
\multiput(\hA,0)(2,0)\hR{\multiput(0,0)(0,2)\Np{\circle*{0.5}}}
}}
\end{picture}

%% file: macroStaggerPatches.tex
\setlength{\unitlength}{1.1ex}
\ifcase2 
   \def\N{1}\def\Nm{0}\def\Np{2}\def\NN{2}\def\NNp{3}
\or\def\N{2}\def\Nm{1}\def\Np{3}\def\NN{4}\def\NNp{5}
\or\def\N{3}\def\Nm{2}\def\Np{4}\def\NN{6}\def\NNp{7}
\or\def\N{5}\def\Nm{4}\def\Np{6}\def\NN{10}\def\NNp{11}
\fi%
\begin{picture}(71,48)
\put(1,2.5){\vector(1,0){68.5}\put(0.5,-1){$x$}}
\put(2.5,1){\vector(0,1){45}\put(-1,45.7){$y$}}
\setcounter{i}{0}
\multiput(6.8,2)(12,0)6{\stepcounter{i}\line(0,1){1}\put(-0.9,-1.6){$X_{\arabic{i}}$}}
\setcounter{i}{0}
\multiput(2,7)(0,12)4{\stepcounter{i}\line(1,0){1}\put(-3,-1){$Y_{\arabic{i}}$}}
\put(3,3){
\setcounter{j}{0}
\multiput(0,1)(0,12)4{\stepcounter{j}
\setcounter{i}{0}
\multiput(0,0)(12,0)6{\stepcounter{i}
\color{green}
\ifodd\value{i}%
\multiput(0,0)(0,1){\NNp}{\line(1,0){\NN}}%
\multiput(0,0)(1,0){\NNp}{\line(0,1){\NN}}%
\else\ifodd\value{j}%
\multiput(0,0)(0,1){\NNp}{\line(1,0){\NN}}%
\multiput(0,0)(1,0){\NNp}{\line(0,1){\NN}}%
\fi\fi
\def\cc{white}\def\cx{yellow}\def\cy{yellow}%
\def\hb{1}\def\Nb{\N}
\ifodd\value{j}%
  \ifodd\value{i}
    \def\cc{black}\def\cx{cyan}\def\cy{red}%
    \def\ha{1}\def\hr{\N}%
    \def\hA{1}\def\hR{\N}%
  \else
    \def\cc{cyan}\def\cx{black}\def\cy{red}%
    \def\ha{1}\def\hr{\N}%
    \def\hA{2}\def\hR{\Nm}
  \fi%
\else%
  \ifodd\value{i}
    \def\cc{red}\def\cx{cyan}\def\cy{black}%
    \def\ha{2}\def\hr{\Nm}
    \def\hA{1}\def\hR{\N}%
  \else
    \def\ha{0}\def\hr{0}\def\hA{0}\def\hR{0}
  \fi%
\fi%
\color{\cc}
\multiput(\hb,\hb)(2,0)\Nb{\multiput(0,0)(0,2)\Nb{\circle*{0.5}}}
\color{\cx}
\multiput(0,\ha)(2,0)\Np{\multiput(0,0)(0,2)\hr{\circle*{0.5}}}
\color{\cy}
\multiput(\hA,0)(2,0)\hR{\multiput(0,0)(0,2)\Np{\circle*{0.5}}}
}}
}
\end{picture}

%% file: macroStaggerCell.tex
\setlength{\unitlength}{1.1ex}
\ifcase2 
   \def\N{1}\def\Nm{0}\def\Np{2}\def\NN{2}\def\NNp{3}
\or\def\N{2}\def\Nm{1}\def\Np{3}\def\NN{4}\def\NNp{5}
\or\def\N{3}\def\Nm{2}\def\Np{4}\def\NN{6}\def\NNp{7}
\or\def\N{5}\def\Nm{4}\def\Np{6}\def\NN{10}\def\NNp{11}
\fi%
\begin{picture}(19,19)
\setcounter{j}{0}
\multiput(0,1)(0,12)2{\stepcounter{j}
\setcounter{i}{0}
\multiput(0,0)(12,0)2{\stepcounter{i}
\color{green}
\ifodd\value{i}%
\multiput(0,0)(0,1){\NNp}{\line(1,0){\NN}}%
\multiput(0,0)(1,0){\NNp}{\line(0,1){\NN}}%
\else\ifodd\value{j}%
\multiput(0,0)(0,1){\NNp}{\line(1,0){\NN}}%
\multiput(0,0)(1,0){\NNp}{\line(0,1){\NN}}%
\fi\fi
\def\cc{white}\def\cx{yellow}\def\cy{yellow}%
\ifodd\value{j}%
  \ifodd\value{i}
    \def\cc{black}\def\cx{cyan}\def\cy{red}%
    \def\ha{1}\def\hr{\N}%
    \def\hA{1}\def\hR{\N}%
  \else
    \def\cc{cyan}\def\cx{black}\def\cy{red}%
    \def\ha{1}\def\hr{\N}%
    \def\hA{2}\def\hR{\Nm}
  \fi%
\else%
  \ifodd\value{i}
    \def\cc{red}\def\cx{cyan}\def\cy{black}%
    \def\ha{2}\def\hr{\Nm}
    \def\hA{1}\def\hR{\N}%
  \else
    \def\ha{0}\def\hr{0}%
    \def\hA{0}\def\hR{0}%
  \fi%
\fi%
\color{\cc}
\multiput(1,1)(2,0)\N{\multiput(0,0)(0,2)\N{\circle*{0.5}}}
\color{\cx}
\multiput(0,\ha)(2,0)\Np{\multiput(0,0)(0,2)\hr{\circle*{0.5}}}
\color{\cy}
\multiput(\hA,0)(2,0)\hR{\multiput(0,0)(0,2)\Np{\circle*{0.5}}}
}}
\end{picture}

%% file: macroStaggerPatchesX.tex
\setlength{\unitlength}{1ex}
\ifcase4 
\or\def\N{1}\def\Nm{0}\def\Np{2}\def\NN{2}\def\NNp{3}
\or\def\N{2}\def\Nm{1}\def\Np{3}\def\NN{4}\def\NNp{5}
\or\def\N{3}\def\Nm{2}\def\Np{4}\def\NN{6}\def\NNp{7}
\or\def\N{4}\def\Nm{3}\def\Np{5}\def\NN{8}\def\NNp{9}
\or\def\N{5}\def\Nm{4}\def\Np{6}\def\NN{10}\def\NNp{11}
\fi%
\begin{picture}(71,48)
\put(1,2.5){\vector(1,0){68.5}\put(0.5,-1){$x$}}
\put(2,1){\vector(0,1){45}\put(-1,45.7){$y$}}
\setcounter{i}{0} \put(2.8,0){
\multiput(\N,2)(12,0)6{\stepcounter{i}\line(0,1){1}\put(-0.9,-1.6){$X_{\arabic{i}}$}}}
\setcounter{i}{0} \put(0,3){
\multiput(1.5,\N)(0,12)4{\stepcounter{i}\line(1,0){1}\put(-3,-1){$Y_{\arabic{i}}$}}}
\put(2,2){
\setcounter{j}{0}
\multiput(0,1)(0,12)4{\stepcounter{j}
\setcounter{i}{0}
\multiput(0,0)(12,0)6{\stepcounter{i}
\color{green}
\ifodd\value{i}%
\multiput(0,0)(0,1){\NNp}{\line(1,0){\NN}}%
\multiput(0,0)(1,0){\NNp}{\line(0,1){\NN}}%
\else\ifodd\value{j}%
\multiput(0,0)(0,1){\NNp}{\line(1,0){\NN}}%
\multiput(0,0)(1,0){\NNp}{\line(0,1){\NN}}%
\fi\fi
\ifodd\value{j}%
  \ifodd\value{i}
    \color{black}
    \multiput(0,0)(2,0)\Np{\multiput(0,0)(0,2)\Np{\circle*{0.5}}}
    \color{red}
    \multiput(0,1)(2,0)\Np{\multiput(0,0)(0,2)\N{\circle*{0.5}}}
    \color{cyan}
    \multiput(1,0)(2,0)\N{\multiput(0,0)(0,2)\Np{\circle*{0.5}}}
  \else
    \color{cyan}
    \multiput(0,0)(2,0)\Np{\multiput(0,0)(0,2)\Np{\circle*{0.5}}}
    \color{red}
    \multiput(1,1)(2,0)\N{\multiput(0,0)(0,2)\N{\circle*{0.5}}}
    \color{black}
    \multiput(1,0)(2,0)\N{\multiput(0,0)(0,2)\Np{\circle*{0.5}}}
  \fi%
\else%
  \ifodd\value{i}
    \color{red}
    \multiput(0,0)(2,0)\Np{\multiput(0,0)(0,2)\Np{\circle*{0.5}}}
    \color{black}
    \multiput(0,1)(2,0)\Np{\multiput(0,0)(0,2)\N{\circle*{0.5}}}
    \color{cyan}
    \multiput(1,1)(2,0)\N{\multiput(0,0)(0,2)\N{\circle*{0.5}}}
  \else
  \fi%
\fi%
}}
}
\end{picture}